\documentclass[prd,letterpaper,onecolumn,amsmath,amsfonts,amssymb,nofootinbib]{revtex4}
\pdfoutput=1
\usepackage {amsmath}  
\usepackage[dvips]{graphicx}              
\usepackage{float}           
\usepackage{feynmp}        
\usepackage{amssymb}                                  
\usepackage[colorlinks=true                                                                       
,urlcolor=blue                
,anchorcolor=blue                    
,citecolor=blue         
,filecolor=blue            
,linkcolor=blue      
,menucolor=blue       
,pagecolor=blue  
,linktocpage=true
,pdfproducer=medialab                             
,pdfa=true            
]{hyperref}  
 \usepackage{csquotes}      
\newcommand{\be}{\begin{equation}}
\newcommand{\ee}{\end{equation}}
\newcommand{\bear}{\begin{eqnarray}} 
\newcommand{\eear}{\end{eqnarray}}

\newcommand{\lapproxeq}{\lower .7ex\hbox{$\;\stackrel{\textstyle
<}{\sim}\;$}}
\newcommand{\gapproxeq}{\lower .7ex\hbox{$\;\stackrel{\textstyle
>}{\sim}\;$}}
\newcommand{\stackdown}[2]{\lower 1.4ex\hbox{$\;\stackrel{\textstyle{#1}}
{\scriptstyle{#2}}\;$}}
\newcommand{\beq}{\begin{equation}}
\newcommand{\eeq}{\end{equation}}
\newcommand{\ba}{\begin{eqnarray}}
\newcommand{\ea}{\end{eqnarray}}

\newcommand{\bea}{\begin{eqnarray}}
\newcommand{\eea}{\end{eqnarray}}

\makeatletter
\def\slash{\@ifnextchar[{\fmsl@sh}{\fmsl@sh[0mu]}}
\def\fmsl@sh[#1]#2{%
  \mathchoice
    {\@fmsl@sh\displaystyle{#1}{#2}}%
    {\@fmsl@sh\textstyle{#1}{#2}}%
    {\@fmsl@sh\scriptstyle{#1}{#2}}%
    {\@fmsl@sh\scriptscriptstyle{#1}{#2}}}
\def\@fmsl@sh#1#2#3{\m@th\ooalign{$\hfil#1\mkern#2/\hfil$\crcr$#1#3$}}
\makeatother
\usepackage{color}

\definecolor{orange}{rgb}{0.9,0.2,0}
\definecolor{brown}{rgb}{0.7,0.3,0.2}
\definecolor{fuxia}{rgb}{1,0,1}
\definecolor{skyblue}{rgb}{0,0.1,0.9}
\definecolor{violetred}{rgb}{0.8,0.13,0.56}
\definecolor{deeppink}{rgb}{1.00,0.08,0.5}
\definecolor{pink}{rgb}{1.00,0.75,0.80}
\definecolor{orchid}{rgb}{0.85,0.44,0.84}
\definecolor{lightpink}{rgb}{1.00,0.71,0.76}
\definecolor{bluish}{rgb}{0,0.6,0.8}  
\begin{document}

 %%%
\title{Issues in  Palatini  ${\cal{R}}^2$ inflation : Bounds on  the Reheating Temperature   } 

\author{A. B. Lahanas$^{a}$ }
\email{alahanas@phys.uoa.gr}

\vspace{0.1cm}
\affiliation{
${}^a$ National and Kapodistrian University of Athens, Department of Physics,
Nuclear and Particle Physics Section,   GR--157 71   Athens,~Greece 
\\
} 
  
%%% 
\vspace*{2cm}    
\begin{abstract}
We consider ${\cal{R}}^2$-inflation in  Palatini gravity, in the presence of scalar fields coupled to gravity. These theories,  in the Einstein frame, and  for one scalar  field $h$, share common features with $K$ - inflation models.  We apply this formalism for the study of single-field inflationary models, whose  potentials  are monomials, 
$ V \sim h^{n} $,  with $ n $  a positive  even integer. We also study the Higgs model  non-minimally coupled to gravity.  
With  ${\cal{R}}^2$-terms coupled to gravity as $\sim \alpha {\cal{R}}^2  $, with $\alpha$ constant, the instantaneous reheating  temperature $T_{ins}$, is bounded by $ T_{ins} \leq   { 0.290 \,   m_{Planck}} / {\,  \alpha^{1/4}} $, with the upper bound being saturated for large $\alpha$. For such large $\alpha$ need go beyond slow-roll to calculate reliably the cosmological parameters, among these the  end of inflation through which  $T_{ins}$ is determined. In fact, as inflaton rolls towards the end of inflation point, the quartic in the velocity terms, unavoidable in Palatini gravity, play a significant role and can not be ignored. The values of $\alpha$, and other parameters, are constrained by cosmological data, setting  bounds on the inflationary scale 
$M_{s} \sim 1/\sqrt{\alpha}$ and  the reheating temperature of the Universe.
\end{abstract} 
%%%
%%%  
\maketitle
 

\section{Introduction} 

The Palatini formulation of General Relativity (GR), or first-order formalism, is an alternative to the well-known  metric formulation, or  second-order formalism.  In the latter the space time connection is determined by the metric while in the Palatini approach the connection $\Gamma^\mu_{\lambda \sigma}$ is treated as  an independent variable
\cite{Sotiriou:2006hs,Sotiriou:2006qn,Sotiriou:2008rp,Borunda:2008kf,DeFelice:2010aj,Olmo:2011uz,Capozziello:2011et,Clifton:2011jh,Nojiri:2017ncd}.
It is through  the equations of motion that $\Gamma^\mu_{\lambda \sigma}$ receive the well known form of the Christoffel symbols,  describing thus  a metric connection.  Within the context of GR the two formulations are  equivalent.  However in the presence of fields that are coupled in a non-minimal manner to gravity this no longer holds \cite{Sotiriou:2006hs,Sotiriou:2006qn,Sotiriou:2008rp}, and  the two formulations describe  different physical theories.    

Encompassing the popular inflation models into  Palatini Gravity, in an effort to  describe the cosmological evolution of the Universe, leads to different cosmological predictions, from the metric formulation,  due to the fact that the dynamics of the two approaches differ. A notable example is the Starobinsky model, for instance, where except the graviton there exists an additional propagating scalar degree of freedom,  the scalaron, whose mass is related to the coupling of   the ${\cal{R}}^2$ term. In the Einstein frame this emerges as a dynamical scalar field, the inflaton, moving under the influence of  the celebrated Starobinsky potential, 
\cite{Starobinsky:1980te,Mukhanov:1981xt,Starobinsky:1983zz}.   
Within the framework  of the Palatini Gravity,  in any  $f({\cal{R}})$ theory \cite{Sotiriou:2008rp}, there are no  extra  propagating degrees of freedom, that can  play the role of the inflaton, and hence the inflaton  has to be put in by hand as an additional scalar degree of freedom.   

The differences between metric and Palatini formulation in the cosmological  predictions, as far as inflation is concerned, arise from the non-minimal couplings of the scalars,  that take-up the role of the inflaton. These couplings  are different in the two approaches. This has been first  pointed out in  \cite{Bauer:2008zj} and has attracted the interest of many authors since,  
\cite{Koivisto:2005yc,Tamanini:2010uq,Bauer:2010jg,Enqvist:2011qm,Borowiec:2011wd,Stachowski:2016zio,Fu:2017iqg,Rasanen:2017ivk,Tenkanen:2017jih,Racioppi:2017spw,Markkanen:2017tun,Jarv:2017azx,Rasanen:2018ihz,Racioppi:2018zoy,Carrilho:2018ffi,Enckell:2018kkc,Bombacigno:2018tyw,Enckell:2018hmo,Antoniadis:2018ywb,Antoniadis:2018yfq,Rasanen:2018fom,Almeida:2018oid,Takahashi:2018brt,Kannike:2018zwn,Tenkanen:2019jiq,Shimada:2018lnm,Wu:2018idg,Kozak:2018vlp,Jinno:2018jei,Edery:2019txq,Rubio:2019ypq,Jinno:2019und,Giovannini:2019mgk,Tenkanen:2019xzn,Bostan:2019uvv,Bostan:2019wsd,Tenkanen:2020xbb,Gialamas:2019nly,Racioppi:2019jsp}, 
with still continuing activity,  
 \cite{Tenkanen:2020dge,Lloyd-Stubbs:2020pvx,Tenkanen:2020cvw,Das:2020kff,McDonald:2020lpz,Shaposhnikov:2020fdv,Bekov:2020dww,Enckell:2020lvn,Lykkas:2021vax,Jarv:2020qqm,Karam:2021wzz,Gialamas:2020snr,Karam:2020rpa,Gialamas:2020vto,Karam:2021sno,Gialamas:2021enw,Gialamas:2021rpr,Annala:2021zdt,Racioppi:2021ynx,Giovannini:2021due,Cheong:2021kyc,Mikura:2021clt,Dioguardi:2021fmr,Ito:2021ssc,Racioppi:2021jai,Rigouzzo:2022yan,Dimopoulos:2020pas,Dimopoulos:2022tvn,Dux:2022kuk,Dimopoulos:2022rdp,He:2022xef}.   
 
Measurements of the cosmological parameters, by various collaborations,  has tighten the allowed window of these observables which in turn constrain, or even exclude, particular  inflationary models, \cite{Planck:2018jri,Planck:2018vyg,Ade:2018gkx,BICEP:2021xfz}.  In particular, the  spectral index $n_s$ and the bounds on the tensor-to-scalar ratio $r$ impose severe restrictions and not all models can be compatible with the observational data
{\footnote{ In this work, standard assumptions are made for neutrino masses and their effective number. Relaxing these it induces substantial  shifts in $n_s$  \cite{Gerbino:2016sgw}.
}}.
%%%%%%%%%%
The precise measurements of the primordial scalar perturbations, and of the associated power spectrum amplitude $A_s$,  imply  constrains for the scale of inflation in models encompassed in the framework of the metric or Palatini formulation, which are more stringent in the case of 
Palatini Gravity as has been shown in \cite{Gialamas:2019nly}.
  
 In this work we shall consider ${\cal{R}}^2$ theories, in the framework of the Palatini Gravity, and study the cosmological predictions of some  popular models existing in  literature, with emphasis on the maximal reheating temperature, or instantaneous reheating temperature.  We will show that there are strict theoretical  bounds on it which are saturated  when the  couplings $\alpha$, associated with the 
 ${\cal{R}}^2$-term,   is large.  To this goal need go beyond slow-roll approximation, to extract reliable predictions, since quartic in the velocity terms of the inflaton play a crucial role.
 Assuming instantaneous reheating the cosmological data impose upper bounds on  $\alpha$, or same, lower bound on the inflationary scale, which also hold for lower reheating temperatures. 
  
 This paper is organized as follows :  
 
 In section II, we present the salient features and give the general setup of   $f({\cal{R}})$ - Palatini Gravity
  %%%%
 {\footnote{
 Throughout this paper  the Ricci scalar will be  denoted by   ${\cal{R}}$. 
 }}
 ,
  in the presence of  an arbitrary number of scalar fields, coupled to Palatini Gravity in a non-minimal manner, in general.  Although this is not new, as this effort has been undertaken by other authors, as well,  we think that the general, and model-independent, expressions  we arrive at, are worth being discussed.
 We  focus on  ${\cal{R}}^2$ theories for which the passage to the Einstein frame is analytically implemented. 
 These theories have a gravity sector, specified by two arbitrary functions, sourcing, in general,  non-minimal couplings of the scalars involved in Palatini Gravity, and a third function which is the scalar potential. 
 In the Einstein  frame, and when a single field is present, these models have much in common with the $K$ - inflation models
 \cite{ArmendarizPicon:1999rj}. 
 
 In section III, we discuss the arising background equations of motions and discuss the slow-roll mechanism, paying special  attention to end of inflation and its validity within the slow-roll scheme. We find that in some cases need go beyond slow-roll to determine the end of inflation which controls  the instantaneous reheating temperature and the cosmological parameters.  

In section IV we discuss various aspects of the inflationary evolution of these models, in the general case, and extract useful conclusions, which hold even when the evolution of inflaton, as it approaches the minimum of the scalar potential, deviates significantly from  slow-roll. 

Section V deals with the instantaneous reheating temperature and its bounds set on it which are dictated by the pertinent backrground equations. Strict upper bounds are derived which are saturated when the parameter $\alpha$, defining the  coupling of the 
$ {\cal{ R }}^2$-terms to gravity,  is large.  These could not have been predicted within  the  slow-roll  scheme. 
Moreover,  assuming that reheating  is instantaneous, we explore the bounds set by the cosmological observables, on the parameters of  particular inflation models, namely the class of models in which the scalar field $h$, is characterized by monomial potentials  $\sim h^{n}$, with $ n $ a positive  even integer, and the Higgs model.  
The power spectrum amplitude $A_s$ results to fine tuning of the parameters of the potential,  while the spectral index   $n_s$ and the tensor to scalar ratio $r$, set bounds on  $\alpha$, and therefore  bounds on the inflation scale and the  instantaneous reheating temperature,  $T_{ins}$.  The latter can be as large as 
$ \sim  10^{15} \, GeV$, the larger  values attained for the  smaller allowed  value of the parameter $\alpha$.
  
 In sections VI,  we end up with our conclusions.

\section{The model}

In this section we shall outline the general setup, and follow the methodology and notation used in 
\cite{Gialamas:2019nly}. More details, if needed, can be found in this reference. 
The starting point is  an action involving scalar fields $h^J$ which are coupled to Palatini  gravity in the following manner, 
\bea
S \, = \, \int \, d^4 x \, \sqrt{ - g} \, \left( \, f({\cal{R}}, h) + \dfrac{1}{2} G_{IJ}(h) \, \partial h^I \partial h^J - V(h)  \, \right)\, .
\label{act1}
\eea 
In it ${\cal{R}}$ is the scalar curvature, in the Palatini formalism, and $f({\cal{R}}, h)$ an arbitrary function of the scalars $h^J$ and 
${\cal{R}}$.   Following standard procedure we write this action in the following manner,
introducing an auxiliary field $\Phi$,  
\bea
S \, = \, \int \, d^4 x \, \sqrt{ - g} \,  \left( \, f(\Phi, h) + f^\prime(\Phi, h) \, (  {\cal{R}} - \Phi) + \dfrac{1}{2} G_{IJ}(h) \, \partial h^I \partial h^J - V(h) \, \right) \, .
\label{act2}
\eea
In this $  f^\prime(\Phi, h)$ denotes the derivative with respect $\Phi$. 
This action can be written as follows, in  Jordan frame,
\bea
S \, = \, \int \, d^4 x \, \sqrt{ - g} \,  \left( \,  \;
\psi \, {\cal{R}} +  \dfrac{1}{2} G_{IJ}(h) \, \partial h^I \partial h^J  - \psi \Phi + f(\Phi, h) - V(h) \,  \right) \, ,
\label{act3}
\eea
where  $\psi$ in defined  by, 
\bea
\psi = \dfrac{ \partial f ( \Phi, h)}{ \partial \Phi} , \quad \text{with \, inverse} \quad \Phi = \Phi ( \psi, h)   \, .
\label{psiphi}
\eea 
%%%
One can go to the Einstein frame by performing a Weyl transformation of the metric
\bea
% g_{\mu \nu} = \lambda \, {\bar g}_{\mu \nu} , \quad  \text{with} \quad \lambda \, \psi=\dfrac{1}{2} \, ,
g_{\mu \nu} =  \, { {\bar g}_{\mu \nu} } \,  /  \, { 2 \psi } 
\eea 
and that done the theory   receives the following form,
\bea
S \, = \, \int \, d^4 x \, \sqrt{ - \bar{g}} \, \,   \left( \;
 \, \dfrac{\overline{\cal{R}}}{2} +  \dfrac{1}{4  \psi} \, G_{IJ}(h) \, \partial h^I \partial h^J 
 - \dfrac{1}{4  \psi^2} \,  ( \psi \Phi - \, f(\Phi, h) + V(h))  \, \right) \,  .
\label{ein1}
\eea
We can further eliminate the field $\psi$, using its equation of motion, 
\bea
\psi \, (\partial h )^2 \, = \, \psi \, \ \Phi   - 2 \, f(\Phi, h)  + 2 V(h) \, , 
\label{solve1}
\eea
where, in order  to speed up notation,  we have  denoted $ G_{IJ}(h) \, \partial h^I \partial h^J =  (\partial h )^2$. 
Note that (\ref{solve1})  is not solvable, in general, however in ${\cal{R}}^2$-theories this is feasible.  

In the following we shall focus on such theories, with a single field $h$ present, with  
 $ f(h , {\cal{R}})$  quadratic in the curvature, having therefore the form
\bea
f({\cal{R}}, h) \, = \, \dfrac{g(h)}{2} \,  {\cal{R}} \, + \,\dfrac{  {\cal{R}}^2}{ 12 M^2(h) } \,   .
\label{fr}
\eea
Since a single scalar field is assumed its kinetic term can be always brought to the form $ \, (\partial h )^2  / 2 $, that is in the action  (\ref{act1})  the field can be taken  canonically normalized.  Therefore in this theory there are three arbitrary functions, namely $g(h), M^2(h), V(h)  $, and any choice of them specifies a particular model. We have set the reduced Planck mass 
$ m_{Planck} \equiv m_P = (8 \pi G_N )^{-1/2}$ dimensionless and equal to unity and thus  all quantities in (\ref{fr}) are dimensionless. When we reinstate dimensions the functions $g, V$ have dimensions $mass^2, mass^4$, respectively, while $M^2$ is dimensionless. 
Note that a non-trivial field dependence of the functions $g(h)$, and/or $ M^2(h)$, is a manifestation of non-minimal coupling of the scalar $h$ to Palatini Gravity.
We recall that in  Palatini formalism  there is no a scalaron field,  associated with an additional propagating degree of freedom,  which in the Einstein frame of the metric formulation  plays the role of the inflaton.

With the function $ f({\cal{R}}, h) $, as given by (\ref{fr}), we get from  Eq. (\ref{psiphi}), 
\bea
\psi = \dfrac{g(h)}{2} \, + \, \dfrac{\Phi}{ 6 M^2(h) } \, , 
\label{ppp1}
\eea 
%% whose inverse is, 
%% \bea
%% \Phi = 6 M^2(h)  \, \left( \psi - \dfrac{ g(h)}{ 2}  \right) \, . 
%% \label{sigma}
%% \eea
and   (\ref{solve1}) is solved  for $\psi$  in a trivial manner,  yielding
\bea
\psi = \dfrac{4 \, V + 3 M^2 g^2 }{ 2 (\partial h )^2  + 6 M^2 g }  \, .
\label{psisol}
\eea
In this way  $\psi$, an hence $\Phi$, from (\ref{ppp1}), are expressed in terms of $ h, (\partial h )^2 $. 
Plugging $\psi, \Phi$ into (\ref{ein1}) we get, in a straightforward manner
\bea
S \, = \, \int \, d^4 x \, \sqrt{ - {g}} \, \,  \left(  \;
 \, \dfrac{{\cal{R}}}{2} +  \dfrac{K(h)}{2} \,  (\partial  h )^2 \, + \, \dfrac{L(h)}{4} \,  (\partial h )^4 \, - \, U_{eff}(h) \, \right) \,  .
\label{einfinal}
\eea 
In this action we have suppressed the bar in the  scalar curvature and also  in $ \sqrt{ -g}$, and in order to simplify notation   we have denoted $  \partial_\mu h \partial^\mu h$  by $  (\partial h )^2$ and 
$(  \partial_\mu h \partial^\mu h)^2$ by $ (\partial h )^4 $.  
Note the appearance of quartic terms $  (\partial h )^4$ in the action. As for the functions $K,L, U_{eff}$, appearing in (\ref{einfinal}), they  are analytically given by
\bea
L(h) \, = \, ( 3 M^2 g^2 + 4 V)^{-1} \, , \, K(h)  \, = \, 3 M^2 g L \, , \, U_{eff} = 3 M^2 V L  \, .
\label{functs}      
\eea
Observe that since terms  up to  ${\cal{R}}^2$  have been considered,   in  $ f({\cal{R}}, h) $ ,   higher than $  (\partial  h )^4$  terms do not appear in the action (\ref{einfinal}). 

 The above Lagrangean may  feature, under conditions,  K - inflation models \cite{ArmendarizPicon:1999rj}, which involve a single field,  described by an  action whose general form is
\bea
S \, = \, \int \sqrt{-g} \;  \left(  \, \dfrac{{\cal{R}}}{2} + p (h, X) \, \right) \, d^4 x   \, .
\label{ssss}
\eea
where $X \equiv (1/2) \partial_\mu h \partial^\mu h$. 
The cosmological perturbations of such models were considered in \cite{Garriga:1999vw} and 
 the importance of a time-dependent speed of sound $c_s$ in K - inflation models was emphasized in \cite{Lorenz:2008je} and   cosmological constraints were derived, where improved expressions for the density perturbations power spectra were used.  
Specific  models were also considered  in  \cite{Li:2012vta}. 
See also \cite{Nojiri:2019dqc,Odintsov:2019ahz,Mikura:2021ldx,Pareek:2021lxz,Odintsov:2021lum,Oikonomou:2021edm}, for more recent works on these  models, in various contexts.

%%%

In a flat Robertson-Walker metric, where the background field $h$ is only time dependent, the energy density and pressure are given by
\bea
\rho(h, X) \, = \, {K(h)} X + 3 \,L(h) X^2 + U_{eff}(h)
\quad , \quad 
p(h, X) \, = \, {K(h)} X +  \,L(h) X^2 - U_{eff}(h)  \, ,
\label{enpre}
\eea
with $X$ being, in this case, half of  the velocity squared,   $ X = {\dot h}^2 / 2$. 
  
We shall assume that the function $L(h)$ is always positive to avoid  phantoms, which may lead to  an equation of state with  
$w < -1$. This may occur  when $ L < 0$ and $ X $  becomes sufficiently  large. However, there is no restriction on  the sign of $K(h)$ which  may be negative in some regions of the field space, signaling that the kinetic term has the wrong sign in those regions.  Obviously  the sign of $K(h)$ should be positive at the minimum of the potential.  Options where $K$ is negative in some regions, although interesting,  will not  be pursued in this work.   Besides,  we shall assume that the potential is positive $U_{eff}(h) \geq 0$ and bears   a Minkowski vacuum. This ensures that the  energy density is positive definite even when the velocity is vanishing.  
The location of the Minkowski vacuum can be taken to be at $ h = 0$, without loss of generality, by merely shifting appropriately the field $h$. Then having a positive definite potential which vanishes at $h=0$ entails $U_{eff}(0) = 0$ and also 
$ U_{eff}^\prime(0) =0 $. 
When inflation models are considered,  the inflaton  will roll down towards  this minimum signaling the end of inflation and beginning of Universe thermalization. 

%%%%%
Concerning  the potential $U_{eff}$,  appearing in the Lagrangian (\ref{einfinal}) in the Einstein frame, using Eq.  (\ref{functs})  it is trivially shown that it can be cast in the following form
{\footnote{
The quantity $R$ should not be confused with the Ricci scalar ${\cal{R}}  $.
}}
,
\bea
 U_{eff} =   \dfrac{1}{4} \, \left(      {3 \, M^2 } \,  - \dfrac{1}{ R}   \right) =
 \dfrac{3 M^2}{4} \, \left(   1  \,  - g K   \right)
 \quad \text{where} \quad R = \dfrac{ L}{ K^2 } \quad .
%%  \label{potspex}
 \label{potrrr}
\eea
%%%
The two forms of the potential above are equivalent, if the relation $  K = 3 M^2 g L $ of Eq.  (\ref{functs}) is used.
The quantity $R$ appearing in this equation may play an important role, as we shall see, in inflationary evolution.  From (\ref{potrrr}) we see that positivity of  $U_{eff} \geq 0$ entails to having $ R^{-1}  \leq  { 3 M^2 } $. In terms of the potential $V(h)$ appearing in the  action (\ref{act1})  this simply  reads  $ V \geq 0$, as can be seen from the last of Eqs.  (\ref{functs}) . 
Dealing with   positive definite potentials, an upper bound is then established,
\bea
U_{eff} \leq   \dfrac{3 M^2}{4 }   \, .
\label{bbb}
\eea
%%%
as is evident from (\ref{potrrr}).  
Although not necessary, 
this upper bound can be easily saturated, for large $h$, by choosing appropriately the functions involved.  Actually the asymptotic behavior of these functions, for large $h$, control the behavior of the potential in this regime. 
Choosing for  instance the function $R$ to increase, as $h$ becomes large,  then saturation of the above bound is easily  obtained
 If, moreover,   we opt that the function $M^2$  approaches or even be  a constant, for large  $h$-values,  
 while   the function  $R$ increases in this regime,  then the potential reaches a plateau which may yield enough inflation. 
 This is a rather  plausible scenario, which may drive successful inflation, and is obtainable under rather mild assumptions.
However,  other less obvious choices may be available. 

The models studied in this work can be, in general, classified in three main  categories :

\begin{itemize}
\item
%% \vspace{3mm}
\noindent
Models with $ g , \, M^2 = constants $, named  {\bf{M1}} for short for future reference.

These are dubbed {\bf{minimally coupled models}}. In this case the constant $g$ can be taken equal to unity without loss of generality. 
This is implemented by rescaling the metric as $ g_{\mu \nu} \rightarrow g_0^{-1} \, g_{\mu \nu} $, where $g_0 = g$,   accompanied by a redefinition of the field $h \rightarrow g_0^{1/2} \, h $  in the action (\ref{act1}),  with $f({\cal{R}}, h) $ as given by (\ref{fr}) , before going to Einstein frame.
 
\item
\vspace{1mm}
\noindent
Models with  $  M^2 = constant$ and $g$ a function of $h$. These we  name  {\bf{M2}} for short.

In this class of models $g$ can be a function of the field $h$, $g(h)$.  In this case we can  take $g(0) = 1$ by rescaling the metric  and the field $h$, in the way   described previously for the M1 models,  with $g_0$ identified with $g(0)$ . In both cases, M1 or M2, it is tacitly assumed that $g_0 > 0$.

\item
Models with both  $ g , \;  M^2 $  functions of $h$.  These we shall name  {\bf{M3}}.

\end{itemize}

The models $M1, M2$  cover a broad range of interesting models,  studied in the past in various contexts, and shall concern us most. 
Models belonging to the class $M_3$ have been studied in \cite{Lykkas:2021vax}.

As we discussed,
we shall be interested in models with positive semi-definite potential having a single Minkowski vacuum at a point, which without loss of generality we can take it to be located  at $ h = 0$.  Then,  besides $ U_{eff} >0 $, we must  have $ \dfrac{d U_{eff}}{dh } >0  $ for 
$ h >0$, with the sign of the derivative  reversed when $ h < 0$.  
These imply restrictions for the functions describing the aforementioned models. For instance, 
for the minimally coupled models this entails $ \dfrac{dK}{dh } < 0 \; ( > 0)  $ for $ h >0 \; ( h < 0 )$, using Eq. (\ref{potrrr}) and the fact that $ g > 0$. The above requirements  are rather mild and can be  easily satisfied. Therefore many options are available for scalar potentials bearing the characteristics demanded for successful inflation to be possibly  implemented. This will be exemplified in specific models, to be discussed later.

Concluding this section,  we presented a general, and  model independent, framework of ${\cal{R}}^2 $ - theories, in the Palatini formulation of Gravity, which may be useful for the study of inflation.  In the Einstein frame these theories may be considered as 
generalizations of $K$-inflation models.  This formalism will be implemented,  for the study of particular inflationary  models.

\section{The equations of motion and the slow-roll}

 \subsection{The inflationary equations of motion } 

When non-canonical kinetic terms are present the equations of motions for the would be inflaton scalar field $h$ differ from their standard form. As a result, the  cosmological parameters describing the  slow-roll evolution should be modified appropriately. Certainly one can normalize the kinetic term of the scalar field appropriately but this is not always very convenient. Actually  the integrations  needed, in order  to pass from the non-canonical to a canonically normalized  field,  in most of the cases, cannot be carried out analytically. Therefore it proves easier, in certain cases,  to work directly with the non-canonical fields and express the pertinent  cosmological observables in a manner that is appropriate for this treatment.

It is not hard to see that the field $h$ satisfies the equation of motion given by
\bea
( K + 3 L \, \dot{h}^2 ) \ddot{h} +3 H(  K +  L \, \dot{h}^2  ) \, \dot{h} + U^\prime_{eff}(h) 
+\dfrac{1}{4} \, ( 2 K^\prime + 3 L^\prime \, \dot{h}^2 ) \,\dot{h}^2 =0  \, ,
\label{eomun}
\eea
where dots denote derivatives with respect time. If the field were canonical, $K = 1$,  and there were no quartic in the velocity terms, that is $L = 0$,  the equation above receives a much simpler form.  In this, 
the effect of using a non-canonical, in general, field $h$ is encoded in the function $K$. The effect of the presence of terms 
 $({\partial h})^4$ in the  action is encoded within the function $L$.   The terms that depend on $L$ are  multiplied by an extra power of the velocity squared, as compared to the $K$-terms. These cannot be neglected, as we discuss below, since  they are not small in general. 

We can gain more insight if we   use a canonically normalized field, say $\phi$, defined by
 \bea
 \phi = \int_0^h{ \sqrt{K(h)}} \, dh   \, ,
\label{canon}
 \eea
 %%%
 which, however, cannot be always presented  in a closed form, as we have already remarked. 
 The constant of integration has been chosen, without loss of generality, so that $h=0$ corresponds to the value $\phi = 0$ too.
 To avoid ghosts we shall assume that $ K > 0$, so that the integration above makes sense.  Actually if  $K$ is negative the kinetic term of the field $\phi$ would have the wrong sign, i.e.  it would appear as $ - ( {\partial \phi} )^2 $.  It could happen however that this function is negative in some region but at the Minkowski vacuum is strictly  positive.  In this way ghosts are also avoided.  This case, interesting as might be, is not discussed and we prefer to keep a rather conservative view point and take $ K > 0$ in the whole region. 
Then in terms of the field $\phi$  the equation of motion (\ref{eomun}) takes on the form
 \bea
 \left( 1 +    3  \, R \, \dot{\phi}^2 \right) \, \ddot{\phi} +3 H \left(  1 +   R \,  \dot{\phi}^2 \right) \, \dot{\phi} + 
  \dfrac{d U_{eff}}{d \phi} +
 \, \dfrac{3 }{ 4 } \,  \dfrac{d \, R }{ d \phi }  \,
 {\dot{\phi}}^4
 = 0   \, .
 \label{eomnum2}  
 \eea 
 Note the dependence of this equation on the ratio $ R = L / K^2 $ defined earlier in Eq. (\ref{potrrr}). 
  From this form it appears that the smallness of the  $ {\partial h}^4$ terms in the action is quantified by the smallness of the ratio 
  $ \frac{ L}{K^2}  \dot{\phi}^2 \ll 1 $, which is equivalent to  $ \frac{ L}{K}  \dot{h}^2 \ll 1 $. 
  
  The equation governing the evolution as functions of time can be easily converted to differential equation for the velocities as function of the fields.  This is done by noting that there is no explicit dependence on time in either of (\ref{eomun}) or (\ref{eomnum2}).  The velocity of $h$-field, $u(h)$ satisfies
  \bea
 ( K + 3 L \,  u^2 ) \, u \dfrac{du}{dh} +3 H(  K +  L \, u^2  ) \, u + U^\prime_{eff}(h) 
+\dfrac{1}{4} \, ( 2 K^\prime + 3 L^\prime \, u^2 ) \,u^2 =0  \, ,
\label{eomunX} 
  \eea
 which is a first order equation with respect the velocity $u(h)$.  The velocity $\dot{h}$ and the acceleration $\ddot{h}$, as functions of $h$,  in Eq. (\ref{eomunX})    are given by
 \bea
 \dot{h} = u(h) \quad , \quad \ddot{h} = u(h) \dfrac{du(h)}{dh} \quad .
 \label{funith}
 \eea
 This method is well-known in Mathematics
 %%%
 \footnote{
The points at which $u$ vanishes correspond to "cusp" points.  There the acceleration $ \dfrac{du}{dh}  $ becomes infinite. 
There are no cusp-points from the beginning of inflation up to values of the field for which $u$ vanishes for first time. This covers the whole inflation region. 
}
.
%%%
 By the same token, for the normalized inflaton field $\phi$, we have
 \bea
 \left( 1 +    3  \, R \, \upsilon^2 \right) \, \upsilon \frac{d\upsilon}{d\phi} +3 H \left(  1 +   R \,  \upsilon^2 \right) \, \upsilon  + 
  \dfrac{d U_{eff}}{d \phi} +
 \, \dfrac{3 }{ 4 } \,  \dfrac{d \, R }{ d \phi }  \, \upsilon^4 = 0 \quad ,
  \label{eomnum2X}  
 \eea
 where, in this case,  the velocity $\dot{\phi}$ and the acceleration $\ddot{\phi}$,  as functions of $\phi$, are given by
 \bea
 \dot{\phi} = \upsilon(\phi) \quad , \quad \ddot{\phi} = \upsilon(\phi) \dfrac{d\upsilon(\phi)}{d\phi} \quad \quad .
 \label{funitf}
 \eea
Equations (\ref{eomunX}) and (\ref{eomnum2X} ), being first order equations, are more easily solved for values of the fields lying in the inflationary regime.

 %%%%%%%%%%%%
 \subsection{Is slow-roll a valid scenario ?} 
\begin{figure}
\centering
%\begin{minipage}{.7\textwidth} 
  \centering  
    \includegraphics[width=0.45\linewidth]{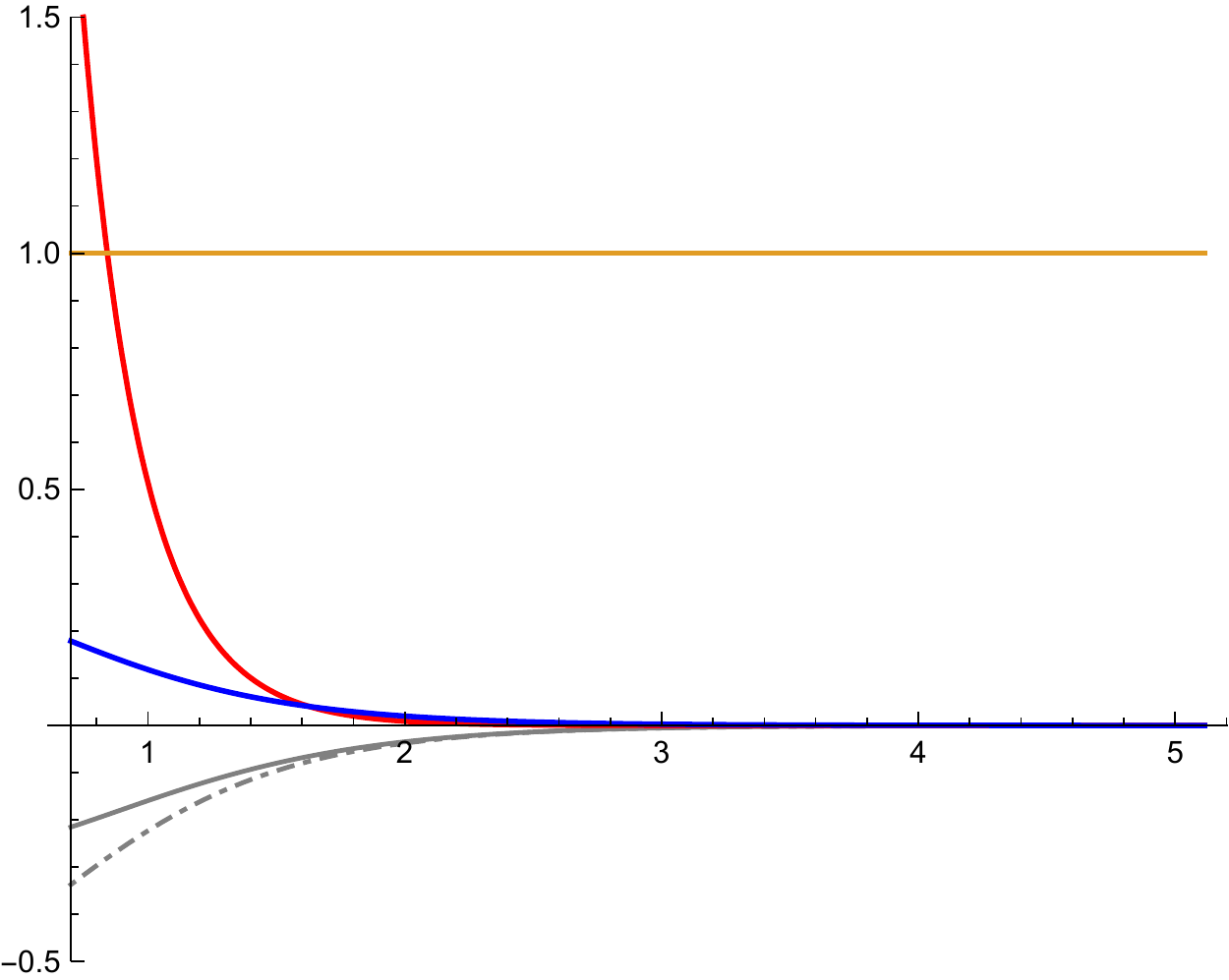}   
   \includegraphics[width=0.45\linewidth]{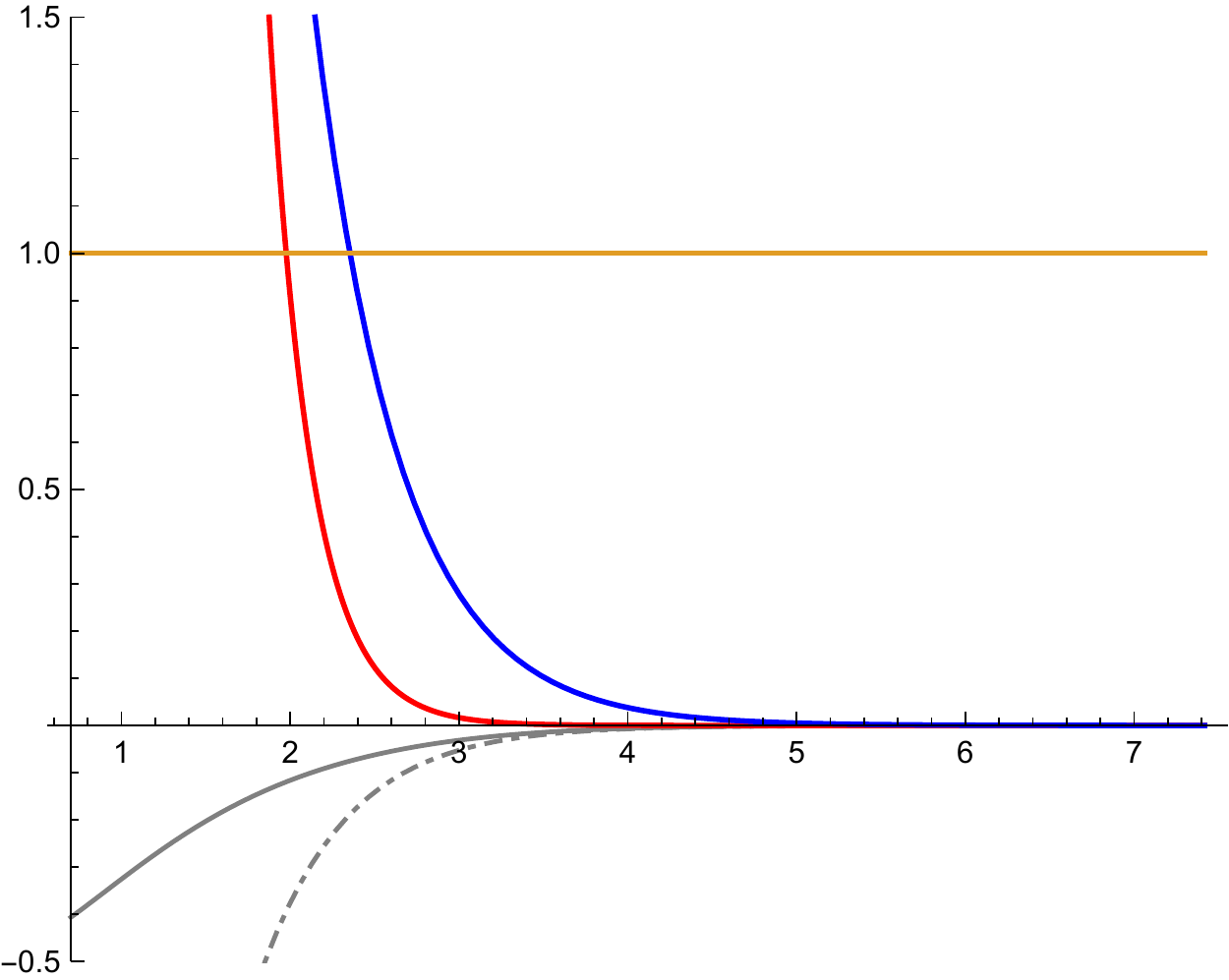} 
%\end{minipage}%
\caption{Evolution of $ \epsilon_V$, red line, and  $R \,  \dot{\phi}^2$, blue line, in the slow-roll approximation in the Model- I . 
as functions of $ \sqrt{c} \phi$.
 On the left, 
 $c = 0.845 $  corresponding  to $ \alpha = 10^{ 10}$ and on the right
$c = 0.845 \times 10^2$  corresponding  to $ \alpha = 10^ {12}$. In both case the parameter $m$ is  $m = 6.5 \times 10^{-6}$. 
   }
   \label{figRR}
\end{figure}

 The usual  slow-roll solution is  not be a valid approximate solution for any values of the parameters involved.  This can be exemplified in certain models,  as we shall see. 
  To be more specific, in the class of models where $M^2 = constant$, which we study in this work, Eq. (\ref{eomnum2})  takes on the form, using Eq (\ref{potrrr}), 
  \bea
 \left( 1 +    3  \, R \, \dot{\phi}^2 \right) \, \ddot{\phi} +3 H \left(  1 +   R \,  \dot{\phi}^2 \right) \, \dot{\phi} + 
  \,( 1 + 3  R^2  \,{\dot{\phi}}^4  ) \,  \dfrac{d U_{eff}}{d \phi}
 = 0   \, .
 \label{eomnum22}  
 \eea 
 %%%
 where the Hubble function is given by, 
 \bea
 3 H^2 = \left( 1 +  \dfrac{3}{2} R \,  \dot{\phi}^2 \right)   \dfrac{ \dot{\phi}^2}{2} + U_{eff}
 \, .
 \label{hubble2}
 \eea
  Neglecting $  R \,  \dot{\phi}^2 $  in the equations above we recover the well-known  form of Friedmann equation for the canonically normalized field $\phi$.  In the regime $  R \,  \dot{\phi}^2 << 1$  slow-roll evolution is  realized. Using the slow-roll expressions   one can  see, in a straightforward manner,  that
  \bea
  R \,  \dot{\phi}^2  = \frac{2 R  \, U_{eff}  }{3}  \, \, \epsilon_V(\phi) \, ,  
  \label{rpara}
  \eea 
  which holds provided $  R \,  \dot{\phi}^2 $ is small enough.   
  In this equation $ \epsilon_V = \frac{1}{2} \, ( {U_{eff}^\prime(\phi)}/{U_{eff}(\phi)} )^2 $.  
  However the smallness of $  \epsilon_V(\phi)  $ does not ensure smallness of $  R \,  \dot{\phi}^2 $.  This is shown in Figure \ref{figRR} for the minimally coupled  model, described  by the  potential  $ V = m^2 h^2 / 2$  and   $ g = 1, M^2 = 1/ 3 \alpha $. 
  In this class of models both $\epsilon_V$ and $ R \,  \dot{\phi}^2 $, depend only on the combination 
  $ c = 2 \alpha m^2$.  We have fixed  $m$ by taking it to be $ m = 6.5 \times 10^{-6}$, or so, suggested by primordial scalar perturbations, as we shall see later. The horizontal axis is  $ \sqrt{c} \phi$. 
  For the case shown on left,  $c = 0.845$, corresponding to $\alpha=10^{10}$,  while on the right $c = 0.845 \times 10^{2}$, corresponding to $\alpha=10^{12}$. Notice the difference in the behaviour of the $  \epsilon_V(\phi)  $ (red solid line) and  $  R \,  \dot{\phi}^2 $ (blue solid line)  functions. For the lower $c$-case, $  R \,  \dot{\phi}^2 $  stays lower than unity all the way up to the point where  $  \epsilon_V= 1 $, marked by the yellow horizontal line.  In this case the usual slow-roll scenario is trusted.  The gray line is the exact solution for the velocity $\upsilon(\phi)$, scaled by $1/\sqrt{a}$,  that is $\upsilon(\phi) / \sqrt{a}$, which has been  derived numerically. The gray dot-dashed line is the corresponding slow-roll approximation.  
  These start deviating significantly, from each other, as soon as $  R \,  \dot{\phi}^2 $ starts becoming sizable.  Note that for the larger $c$-case displayed,  this occurs well  before $  \epsilon_V(\phi)  $ becomes unity, showing that slow-roll  ceases to be a  good approximation although $  \epsilon_V(\phi)  $ is significantly lower than unity.  
 The conclusion is that, for  lower $c$, corresponding to lower $\alpha$,  $  R \,  \dot{\phi}^2 $ is   small, and the slow-roll solution is a good approximation. 
  On the contrary for  large $c$-values, $  R \,  \dot{\phi}^2 $ becomes large  before $  \epsilon_V= 1 $. Therefore   slow-roll in the usual sense can be only realized  in the region of $\phi$-values for which $  R \,  \dot{\phi}^2 $ is small, and thus far from the point where 
  $  \epsilon_V= 1 $.  The evaluation of  the end of inflation, in this case, can be only achieved numerically since approximate slow-roll solutions, based on  $  \epsilon_V= 1 $, cannot be trusted in the region  of $\phi$ for which   $  R \,  \dot{\phi}^2 $ starts approaching unity. 
 
 Concerning the end of inflation, this takes place  when the parameter $\epsilon_1 = - \dot{H}/H^2$ becomes equal to unity. In the slow-roll regime, where 
   $  R \,  \dot{\phi}^2 $ are negligible, $\epsilon_V$ and $\epsilon_1$ almost coincide, assuming de-Sitter expansion,   but this is not the case when 
   $  R \,  \dot{\phi}^2 $ terms start growing and the approximate slow-roll solution no longer holds.  Therefore we rely on $\epsilon_1 = 1$ 
   as the only  reliable means to determine accurately the end of inflation.  This is equivalent to $ \rho + 3 p =0 $,  corresponding to $\ddot{a} = 0$ for the cosmic scale factor, and yields the following relation between the velocity and field value,  at time  $\epsilon_1 $ reaches unity,  
  \bea
\upsilon^2(\phi) \, = \,  \dfrac{2}{3 R} \, \left(  -1 + \sqrt{ 1 + 3 R U_{eff} } \right)
 \label{endofinf}  
  \eea
  In this  $\upsilon(\phi)$ is the velocity $ { \dot{\phi}} $ expressed as function of the field $\phi$, and can be extracted numerically by solving   (\ref{eomnum2X}). 
  In  ordinary inflation models where the  $R$-terms are missing, that is there are no terms quartic in the velocity in the action,  the analog of Eq. (\ref{endofinf}) is  $ \upsilon^2(\phi)    =  U_{eff} $, a well-known result, and this can be solved to yield   the value of the field, $\phi_{end}$ at end of inflation,  if the slow-roll solution is used.  This is actually equivalent to  $\epsilon_V = 1$
 %%%%%
 {\footnote{
When  the inflaton kinetic energy is $ {\dot \phi}^2/2 $,  the condition $\epsilon_1 = 1$ corresponds actually to $\epsilon_H(\phi)  = 1$,  where  $\epsilon_H(\phi), \eta_H(\phi)   $ are slow-roll parameters  defined in   \cite{Liddle:1994dx}. Thus $\phi_{end}$ extracted from  
$\epsilon_V = 1$ can be considered as a first order result.  An  improved value for $\phi_{end}$, using the exact relation 
$\epsilon_H(\phi)  = 1$,   can follow using $  \epsilon_V  = \left(1 + \sqrt{1 - \eta_V/2 }    \right)^2$, \cite{Ellis:2015pla}.  To our knowledge, there  is no such a relation when  $R$-terms. are present, which would approximate (\ref{endofinf}) .
  }}
 %%%%%%%%
  If the $R$-terms are sizable,  before end of inflation,  
  we lack even an approximate solution for $ \upsilon(\phi) $,    and thus 
  (\ref{endofinf}) can be only   tackled numerically,  in order to know  $\phi_{end}$.  In fact using the slow-roll parameter 
  $\epsilon_V$ to determine the  end of inflation  overestimates the value $\phi_{end}$, as this is extracted from   (\ref{endofinf}) , leading to erroneous results concerning cosmological observables, and in particular the energy density at the end of inflation, which determines the reheating temperature of the Universe.

  %%%%%%%%%%%%%% MINI        
%%%%%% MINI   
 
  It should be stressed that our numerical study duly takes  into account the contribution of these terms and no approximation, whatsoever,  is made. We have found numerically  that they can be  indeed  small,  \cite{Enckell:2018hmo,Antoniadis:2018ywb,Antoniadis:2018yfq,Tenkanen:2019jiq}, however this holds in a restricted range of the parameters and is not a general feature. For instance in the non-minimal model discussed previously, with $ V \sim h^2 $,
  the combination $ c = 2 \alpha m^2$   has to be smaller than about unity, for these terms to be small in the entire inflationary region.

\section{Inflationary Evolution - End of inflation    }

The end of inflation is signaled when $\epsilon_1 = 1$, or equivalently when acceleration ends, $\ddot{a} = 0$.  The determination of end of inflation  requires augmented accuracy, at least in some models, as we shall discuss in the sequel. The slow-roll parameter  $\epsilon_1 $ is very small during de Sitter phase and then starts increasing. Eventually becomes equal to unity where Universe acceleration stops. In order to locate when this occurs we find it useful to have an expression for the time derivative of it.   $\epsilon_1 $ can be expressed in the following way,
\bea
\epsilon_1  = \dfrac{3}{2} \dfrac{\rho + p }{\rho} = \dfrac{3}{2} \, \; \dfrac{V^2 + R \, V^4 }{ \rho \; \;} 
\label{veloti}
\quad , \quad 
\eea
where $V$ is the velocity $  V = \dot{\phi}$ and $ R \equiv {L} / {K^2} $ .  Note that the numerator does not explicitly depend on the potential. 
Then the time derivative of $\epsilon_1 $  is of the form,
\bea
\dot{\epsilon}_1 \equiv \dfrac{ d \epsilon_1}{ dt} = \dfrac{ \cal{N} }{ \rho^2}
\label{dereee}
\eea
where the numerator $\cal{N}$ is found to be
\bea
{\cal{N}}= 
2 V \dot{V} \, \left( - \dfrac{R \, V^4}{4} +( \, 1 + 2 \, R \, V^2 ) \, U_{eff}  \right) + 
\left( - \dfrac{V^2}{4} +\, U_{eff}  \right) \, V^5 \, R^\prime - 
( \, 1 +  R  \,V^2 ) \, V^3 \, U^\prime_{eff}
\label{nvalit}
\eea
In this all primed quantities are derivatives with respect $\phi$.

We assume that the evolution  starts from an initial  position $ h > 0$  on the plateau of the potential, or in terms of the canonically normalized inflaton field $\phi > 0$.  Recall the two fields are related by  $ \phi = \int \, \sqrt{K} \, dh  $, and we have chosen   the integration constant so that $\phi = 0$  corresponds to $ h = 0$ too. The  potential is positive definite and exhibits a zero at  $h = 0$ and as a consequence at $\phi = 0$,  if expressed in terms of $\phi$.  Due to its positivity the first derivative of the potential vanishes at the minimum as well.

The inflaton start its journey with a very small, or even  vanishing initial velocity, which soon  becomes negative $ {V} < 0$  and increases in magnitude,  as inflaton rolls towards  the minimum of the potential. At some time its direction is reversed. Therefore there is a time $ t_{acc}$ for which its acceleration vanishes, $\dot{V} ( t_{acc} ) = 0,   $ while $ {V} ( t_{acc} ) < 0  $. Then  $\dot{V}  >  0   $  for  
$ t > t_{acc}$ and thus 
the velocity increases and  eventually at some time $t_1$ it vanishes for the first time, i.e $V(t_1) = 0$. 
Therefore the picture is that $ V < 0$, as long as   $ t < t_1$, attaining its minimum value at $t_{acc}$,  and at the time $t_1$ it vanishes.
At this time the acceleration is still positive, i.e. the velocity continuous being increased passed the time $t_1$. It takes some time after it vanishes, reversing  its direction,  and start oscillating about the minimum of the potential.

From the equation of motion  (\ref{eomnum2})   for the field $\phi$, we have therefore that at this time $t_1$,
\bea
\left(  \dot{V} + \dfrac{ d U_{eff}}{  d \phi} \right) {\biggr\rvert }_{ t_1} \, = \, 0
\eea
However since, as we have discussed, $ \dot{V} (t_1) > 0  $ it follows from the equation above that $  \frac{ d U_{eff}}{  d \phi} {\bigr\rvert }_{ t_1}  < 0 $.  During inflaton's  journey, up to time it reaches the minimum of potential, we have that  
$ \frac{ d U_{eff}}{  d \phi}  \geq 0   $, with the minimum being reached at time $t_0$ and it for later times it becomes negative. Therefore $t_0$ is prior to $t_1$,  $ t_0 < t_1$. That is the minimum of potential, which is located at   $\phi =0$, or $h=0$, is reached before the velocity vanishes for the first time during inflation, which is a rather expected behaviour.

\begin{figure}[H]
\centering
%\begin{minipage}{.7\textwidth} 
  \centering  
  \includegraphics[width=0.8\linewidth]{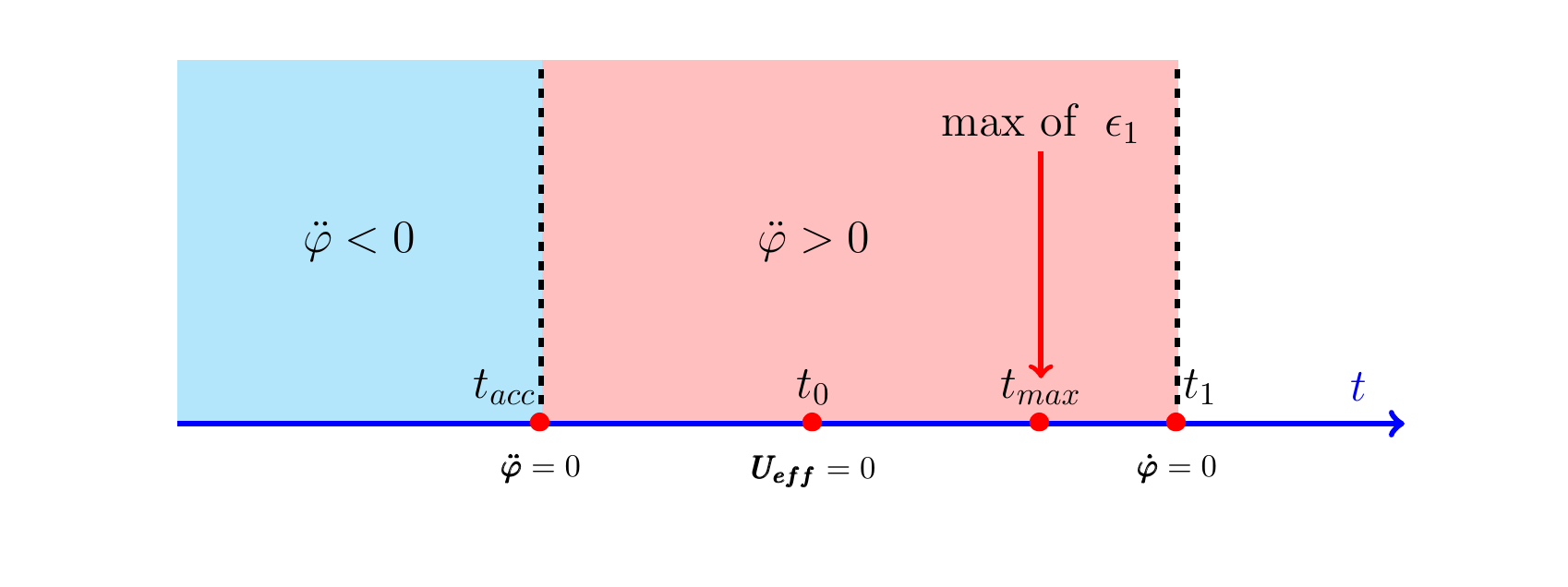}     
%\end{minipage}%
\caption{The ordering of times $ t_{acc}, t_0, t_1 $ is as depicted in this figure for  a wide class of models, including those with $M^2 = constant$. Between $t_{acc}$ and $t_1$ the acceleration 
of inflaton $\phi$ is positive.  
  }
   \label{timeline}
\end{figure}

From the previous discussion we concluded that the time $t_{acc}$, at which the acceleration of $\phi$ vanishes, and hence the velocity $V$ takes its minimum value, is prior to $t_1$.  Its location relative to $t_0$,  the time  inflaton reaches the minimum of the potential, may be derived as follows.
The acceleration is determined from the equations of motion. Using the fact that $U_{eff} , U^\prime_{eff}$ vanish at $t_0$ we have that 
\bea
\dot{V} (t_0) \, = \, 
- \dfrac{ 1}{ 1 + 3 \, R \, V^2 } \; 
\left(  3 \, H \, ( 1 + R \, V^2 ) \, V + \dfrac{3}{4} \; R^\prime  \, V^4  \right) {\biggr\rvert }_{ t_0} 
\quad , \; \; \text{where}  \quad  R^\prime =  \dfrac{d R}{d \phi} 
\label{accit0}
\eea 
Due to the fact that $V(t_0)$ is negative, we have, from this equation, that the acceleration $ \dot{V} (t_0)  $ is positive if 
$  R_0^\prime = R^\prime (t_0) \leq  0  $.  The sign of $ R_0^\prime  $ is related to the derivative of the potential since, 
\bea
\dfrac{ d U_{eff} }{ d \phi } = \,  \dfrac{ 3  }{ 4  } \,  \dfrac{  d M^2 }{  d \phi } + 
\dfrac{1}{4 R^2 }\, \dfrac{ d R }{ d \phi } \quad   .
\label{ddde}
\eea
This conclusion can be drawn using  the form of the potential as given by eq (\ref{potrrr}), which is valid for any $M^2$.  
From this it directly follows, using the fact that  the derivative of the potential vanishes at $\phi = 0$, corresponding to $t_0$,
\bea
R_0^\prime = - 3 \, R_0^2 \, M_0^{2 \; \prime}  \quad  \text{where} \quad  M_0^{2 \; \prime} =  \dfrac{  d M^2 }{  d \phi }  {\biggr\rvert }_{ \phi = 0 } \quad .
\label{R00}
\eea
%%%
Therefore $  R_0^\prime \leq 0  $ in all models with $ M_0^{2 \; \prime}  \geq 0 \, $ and as a consequence   $\dot{V} (t_0) > 0  $. This states that  $t_0$ lies in a region the acceleration is positive, and this occurs after $t_{acc}$. Therefore in models with 
$ M_0^{2 \; \prime}  \geq 0 \, $ we have that  $t_{acc} < t_0$. 
 The  models M1, M2 , which are characterized by  a constant $M^2$, fall within this class  since  $M^{2 \; \prime} = 0 \, $. 
 For models not belonging to this category the condition $ M_0^{2 \; \prime}  \geq 0 \, $ has to be checked as per case. 
 Therefore when  $ M_0^{2 \; \prime}  \geq 0 \, $  we have  $\dot{V} (t_0) > 0  $ and  $t_0$ lies in the region inflaton accelerates before its velocity vanishes for the first time. This occurs for times larger than  $ t_{acc}$.  Concluding,   there is a broad class of models,  those with  $ M_0^{2 \; \prime}  \geq 0 \, $,  for which the time ordering is  $\, t_{acc} < t_0 < t_1 \, $. The situation is shown in Figure \ref{timeline} .
  %%%%%%%%%%%%%% MINI        
%%%%%% MINI 

%%%%%%
Concerning the evolution of $\epsilon_1$, it initially has  a very small positive value and then starts increasing. This vanishes for first time when the velocity does, as is evident from Eq. (\ref{veloti}),  that is at time $t_1$. Therefore it ought develop a maximum at some intermediate time, say $t_{max}$ which is prior to $t_1$, i.e. $ t_{max} < t_1 $.  
Whether $t_{max}$ lies  before or after $t_0$ we do not know as yet. We shall show that in the class of minimally coupled models M1, and also models M2,   $ t_0 < t_{max}$ so that  time ordering  is as 
 $ t_0 < t_{max} < t_1$. That is the minimum of the potential is reached for first time  before $\epsilon_1$ attains its first maximum, and  later, at time $t_1$,   $\epsilon_1$  vanishes for first time along with velocity $V$. 
%%%
 The proof relies on what is the sign of $\cal{N}  $ in Eq. (\ref{nvalit}) at $t_0$. Since the potential and its derivative vanish at $t_0$, due to the fact that  at this time $\phi = h = 0$, the value of $\cal{N}  $ is
 \bea
 {\cal{N}}_0 = {\cal{N}} {\bigr\rvert }_{ t_0} = 
 - \dfrac{ V^5}{ 2 } \; \left( \,  \dot{V} \, R +   \dfrac{R^\prime }{  2} \,   V^2  \right) {\biggr\rvert }_{ t_0} 
 \label{signn}
 \eea
From this it is seen that the sign of $ {\cal{N}}_0  $ follows that of the bracketed quantity  in  equation (\ref{signn}), due to the fact that 
$V(t_0) < 0$.  Recall that velocity is negative for $ t < t_1$ and $t_0$ i.e. earlier than $t_1$ as we have shown before.  Therefore for the sign of $ {\cal{N}}_0  $ we  need study  the  acceleration and also  $R^\prime$ at $t_0$.  For the models M1, M2 the acceleration is positive at the point $t_0$, as we have already shown ( see also Figure \ref{timeline}).  For these models,  on account of Eq. (\ref{ddde}),  $ R^\prime(t_0)   $  vanishes forcing the acceleration   $\dot{V}(t_0)  $  at $t_0$ to be positive, as we have already discussed. Therefore from  (\ref{signn}) we have that  $ {\cal{N}}_0  > 0 $.
This entails that $ \dot{\epsilon}_1(t_0) > 0$, using equation  (\ref{dereee}).  Since $\epsilon_1$ is monotonically increasing until this reaches its maximum at $t_{max}$,  positivity of  $ \dot{\epsilon}_1(t_0) $ states that the maximum of $\epsilon_1$ is reached  later than $t_0$, i.e $ t_0 < t_{max}$,  in the class of models having $M^2 = \text{constant}$, as shown on Figure \ref{timeline}.  
In the same figure the location of $t_{max}$ relative to other critical times in the models M1, M2 is also shown.
As we have stated, 
this covers the class of minimally coupled models, but also non-minimally coupled models in which $g(h)$, designating the coupling of the Ricci term in (\ref{fr}),  is not  constant.

Concerning the maximum value of $\epsilon_1$ cannot exceed $3$. In fact from (\ref{veloti}) we have, writing explicitly the density $\rho$,
\bea
\epsilon_1 \, = \, 3 \, \dfrac{V^2 + R V^4  }{V^2 +( 3 R/  2) V^4 + 2 U_{eff} } `\leq
 3 \, \dfrac{V^2 + R V^4  }{V^2 +( 3 R/  2) V^4 } < 3 
\eea
therefore a strict upper bound on $\epsilon_1$ can be established.  At the time $t_0$ we have, due to the fact that the potential vanishes,
\bea
\epsilon_1 (t_0) = 
3 \, \dfrac{ 1 + R_0 V_0^2  }{1  +( 3 R_0 /  2) V_0^2 } \geq 2 
\eea
where the subscript $0$ means evaluation at $t_0$.  This lower bound on the value of $\epsilon_1(t_0) $ combined with the fact that 
$ \epsilon_1$ is monotonically increasing for all times $ t \leq t_{max}$ ensures that there is certainly  a time $t_{end} < t_0$ at which inflation ends, that is $\epsilon_1(t_{end} ) = 1 $.   
%%
 %%%%%%%%%%%%%% MINI        
\begin{figure}[t]  
\centering
%\begin{minipage}{.7\textwidth} 
  \centering  
  \includegraphics[width=0.55\linewidth]{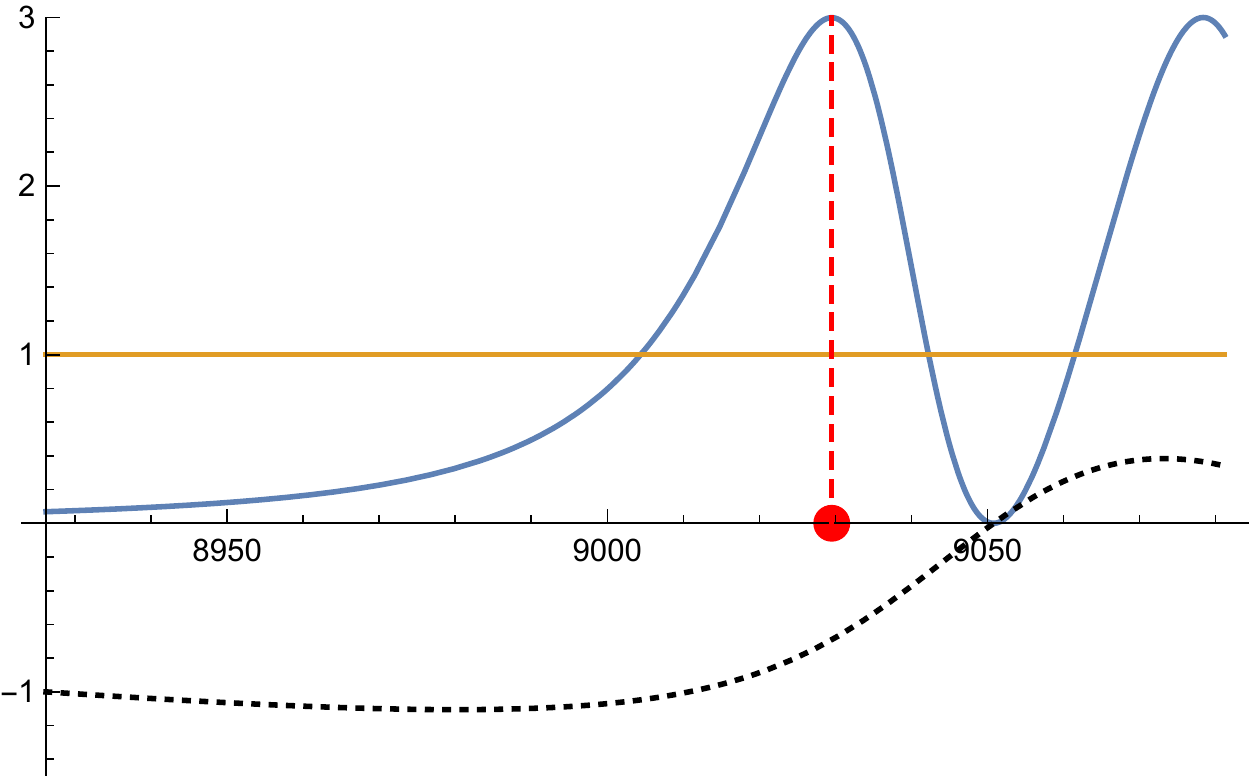}     
%\end{minipage}%
\caption{In model I, and for $ \alpha = 10^8$, we display the evolution of $\epsilon_1$, in blue,   and the velocity  $\dot{\phi}$, dashed line, from some initial $t_{ini}$ time. The horizontal axis  is   the rescaled cosmological time  
$ t / \sqrt{\alpha}$,  and  the velocity  is actually $\dot{\phi}$ divided by its values at $t_{ini}$. 
The horizontal amber line crosses the  $\epsilon_1$ curve at  where   inflation ends, and this is prior to the minimum of the potential, whose location is shown by the red circle. 
The maximum of  $ \epsilon_1$  lies to the right of the minimum, but it is hardly visible on the figure. $\epsilon_1$ vanishes for first time after the minimum is reached. Its  oscillatory behaviour after,   is due to the fact that the velocity  oscillates, after inflaton passes the minimum of potential.
  }
   \label{figinfla}
\end{figure}  
%%%%%% MINI  
 This behaviour is exemplified in Figure \ref{figinfla}  for a model in which $M^2 \equiv  1/ 3 \alpha$ and $g(h) = 1$, and quadratic  potential $ V(h) = \frac{ m^2 \, h^2}{ 2 }$.  The values of the arbitrary parameters are $ \alpha = 10^8$ and 
$ m = 6.5 \times 10^{-5}$.

The  results reached in this section  are useful in our numerical treatment,  in order to locate with the required precision   the end of inflation, in models with $ M^2 = 1/3 \alpha$, especially when the parameter $\alpha$ is large. It is in these cases that inflaton dynamics cannot be described by  slow-roll and the speed of sound, as we shall see, deviates from unity. This is  equivalent to 
having non-negligible contributions from the quartic in the velocity $L$-terms appearing  in the action, as inflaton approaches the end of inflatoin.

\section{Universe's Reheating Temperature}

 The reheating  temperature the Universe reached after its thermalization has been extensively studied and various mechanisms and models have been put under theoretical scrutiny,  \cite{Kofman:1994rk,Shtanov:1994ce,Kofman:1997yn,Chung:1998rq,Kawasaki:1999na,Davidson:2000er,Giudice:2000ex,Martin:2010kz,Allahverdi:2010xz,Podolsky:2005bw,Adshead:2010mc,Mielczarek:2010ag,Easther:2011yq,Dai:2014jja,Harigaya:2013vwa,Munoz:2014eqa,Cook:2015vqa,Gong:2015qha,Rehagen:2015zma,Drewes:2015coa,Cai:2015soa,Ellis:2015pla,deFreitas:2015xxa,Lozanov:2016hid,Lozanov:2017hjm,Dalianis:2016wpu,DiMarco:2018bnw,Hasegawa:2019jsa,He:2018mgb,German:2020iwg,Saha:2020bis,Hamada:2020kuy,He:2020ivk,He:2020qcb,DiMarco:2021xzk,Kawai:2021hvs,Pareek:2021lxz}.  
  Reheating in the framework of the Palatini gravity has been also  studied  in \cite{Rubio:2019ypq,Gialamas:2019nly,Das:2020kff,Karam:2021sno,Cheong:2021kyc}

  As for the number of e-folds left,   $\, N_k = ln \, \dfrac{ a_{end}}{ a }$,   from time  some  scale $k$ crossed the sound horizon to the end of inflation  is given in     \cite{Liddle:2003as,Dodelson:2003vq}.  See also \cite{Martin:2010kz,Lozanov:2017hjm}. 
Note that the dependence on speed of sound $ c_s$ should be  included in  $N_k$,   due to the fact that it may deviate from  unity, as is the case in $K$-inflation models.

The number of e-folds accrued during the  reheating period, 
 $   \Delta N_{reh} $,  is given by  
 \bea
  \Delta N_{reh} \equiv ln \, \dfrac{a_{reh}}{a_{end}} = - \dfrac{1}{3 ( 1+w)} \, ln \, \dfrac{\rho_{reh} }{ \rho_{end }}  \, .
  \label{nfoldreh}
 \eea
 The subscripts  (reh), (end) in the cosmic scale factor and the energy densities denote that these quantities are evaluated at the end of the reheating period and inflation respectively. In terms of Hubble rate $ \rho_{end } = 3 m_P^2 \, H_{end}  $, and therefore this  is known once $t_{end}$, or the values of the fields and their velocities at end of inflation, are known.
  
 The  effective  equation of state parameter $w$ in the reheating period,   is the average
  \bea
 w \, = \, \dfrac{1}{ \Delta N_{reh}} \,  \int_{N_{end}}^{N_{reh}} \; w(N) \, d N  \quad .  
 \label{meanw}  
 \eea 
 The integration variable is the number of e-folds and $  \Delta N_{reh} =  N_{reh} -  N_{end} $, where  
 $  N_{end} , N_{reh}  $,  are the number of e-folds at end of inflation and reheating periods respectively. 
At the end of inflation 
 $w( N_{end} ) = - 1/3$ while $ w(N_{reh}) = 1 / 3$ corresponding to the onset of radiation dominance. 
 Lacking a particular reheating mechanism   the value of $w$ is largely unknown, therefore we shall consider it 
 as a free parameter taking values within some sensible range. 
 In the canonical reheating scenario $w =0$, but  values in the range  $ \simeq 0.0 - 0.25$, or larger, right after inflation,  are also possible in some models
 \cite{Podolsky:2005bw,Lozanov:2016hid,Lozanov:2017hjm,Dux:2022kuk}.
 
In terms of $ \Delta N_{reh}  $,  for given $w$,  one has for the reheating temperature,, see for instance  \cite{Munoz:2014eqa}, 
 \bea
 T_{reh} \, = \, \left(  \dfrac{ 30}{\pi^2 \, } \dfrac{      \, \rho_{end} }{   \, g^{* (reh)} \,} \, \, \right)^{1/4} \,
 exp \left({ - \dfrac{ 3(1+w) \Delta N_{reh}  } {  4}  } \right) \,  .
\label{treh}
 \eea  
 %%%%  
 In our numerical studies we shall adopt the common  values $ g^{* (reh)}  = g_s^{* (reh)} = 106.75 $, corresponding to the SM content, as discussed before, for temperatures above $\sim 1 \, TeV$. 
 %%%%%%
 {\footnote{  
 With 
 $  g^{* (reh)}  = 100 $  Eq. (\ref{treh}) coincides with that given in  \cite{Munoz:2014eqa}.
 }} .
 %%%%%%%
 Note  that since $  {a_{reh}} > {a_{end}} $  we have that $ \Delta N_{reh} \geq 0$, and therefore due to  $w > -1$ the reheating temperature $  T_{reh}$ is bounded from above 
 \bea
 T_{reh} \leq   \left( \dfrac{ 30}{\pi^2 \, } \dfrac{      \, \rho_{end} }{   \, g^{* (reh)} \,} \, \, \right)^{1/4}  \,  .
\label{tins}
 \eea
 The bound on the right hand side  of this  defines  the instantaneous reheating temperature, $  T_{ins}$. The  temperature 
 $  T_{reh}$ reaches this upper  bound when  the reheating process  is instantaneous, in which case $ \Delta N_{reh} = 0$.  Note that 
 for rapid thermalization we have $ \rho_{end} = \rho_{reh} $, from Eq. (\ref{nfoldreh}). 
 The reheating temperature should be  larger than  $  \sim 1 \,  MeV$ so that  Big Bang Nucleosynthesis (BBN) is not upset.  Lower values on $T_{reh}$ have been  established in   \cite{Kawasaki:1999na} and  more recently in  \cite{Hasegawa:2019jsa}   .

The maximum reheating temperature depends, as we shall see, on the value of  of the sound of speed parameter $c_s$ at end of inflation. This is not unity, in general,    due to the fact that in the Palatini formulation of ${\cal{R}}^2$ gravity  higher in the  velocity 
$ \dot{h}$  terms are  unavoidable. In  fact $c_s$ is defined by
\bea
c_s^2 \, = \, \dfrac{ \partial p / \partial X}{ \partial \rho / \partial X}  \, ,
\label{sound}
\eea
where $X$, defined after Eq. (\ref{ssss}), is half the velocity squared.  In terms of the fields $h$, or $\phi$, and their velocities $u$ and $\upsilon$ respectively, this receives   the form
\bea
c_s^2 \, = \, \dfrac{ 1 + L \, u^2  / K}{ 1 + 3 L \, u^2 / K}  \,   
= \, \dfrac{ 1 + R \, \upsilon^2  }{ 1 + 3 R \, \upsilon^2 } 
\label{sound2}
\eea
Inverting this we get,
\bea
\omega \, = \, \dfrac{1 - c_s^2 }{ 3 \, c_s^2 - 1 }
\label{omcs2}
\eea
where $\omega = L \, u^2 /  K$, which is also  equivalent to $\omega   = R \, \upsilon^2$.  Actually $\omega$ is 
the same combination  that appears in the equation of motion for the field $h$, or $\phi$ respectively. 
%%%
The velocity 
$ c_s $ is thus  controlled by $ \omega $, and it is seen   from (\ref{sound2})  that  $c_s^2$  is strictly less than unity and approaches unity only when   $ \omega \ll 1 \,$.  Interestingly enough is also bounded from below by $1/3$. Thus  $ c_s^2$ takes values in the range
\bea
\dfrac{1}{3} <  c_s^2 \; \leq 1 \quad .
\label{boundc}
\eea
These are strict mathematical bounds. The value of  $c_s^2$  at any epoch is known if  one solves the pertinent differential equation for $h$, or equivalently $\phi$.  The upper bound is reached  for vanishing values of $ L \, u^2 /  K$, or $  R \upsilon^2   $,  and the lower limit when 
these quantities get values much larger than unity. As we shall prove the value of $c_s^2  $ when inflation ends determines $\rho_{end}$, and as a consequence the maximum reheating temperature. 

At end of inflation $ \rho + 3 p =0 \;  $ and  the solution of this equation relates the velocity and the position at end of inflation through Eq. (\ref{endofinf}), 
which in terms of the velocity of the field $h$ takes the form
\bea
u^2(h) = \dfrac{2 K}{3 L  } \, 
\left( -1 + \sqrt{1 + \dfrac{ 3 L  U_{eff}}{  K^2}  }\right) \; .
\label{whenend2}
\eea
%%%
Both (\ref{endofinf}), or (\ref{whenend2}),  hold at end of inflation and cannot be treated analytically. 
Only numerically  we can solve the pertinent differential equations and through these determine the end of inflation time, or equivalently the value of the position at end of inflation. Approximate solutions  do exist but they are unreliable, in the present case, 
 due to the presence of the $L$-dependent terms. As we have already discussed, and depending on the values of the parameters involved, the solutions may differ substantially from slow-roll towards end of inflation. 
Therefore   evaluation of end  of inflation period using approximations, relying on  the use of the slow-roll parameters  $ \epsilon_V, \eta_V $,  poorly  determine when the end of inflation actually occurred.  
%%%

Using the expression for the density $\rho$ one can use (\ref{endofinf}), (\ref{whenend2}), to find,
\bea
\rho_{end} \, = \, \frac{ 3}{2 R}   \;  \omega \; ( 1 + \omega ) \; {\biggr\rvert }_{ {end}}  
\label{rendxi}
\eea 
In this,   $\omega$ has been defined previously and in terms of the sound of speed is  given by  Eq. (\ref{omcs2})
{\footnote{
Obviously (\ref{rendxi}) holds provided $\omega $, as defined by equation  (\ref{omcs2}), is non-vanishing. The $\omega=0$ case corresponds to   $R = 0 $,  and hence $c_s^2 = 1$ which is the case in the usual inflation scenarios. In this case it is well-known that 
$ \rho_{end} \, =  \frac{3}{2} \, U_{eff} $ at the end of inflation.
}}
.

Note that all quantities in (\ref{rendxi}) are meant at the end of inflation.
 Combining these we find
\bea
\rho_{end} \, = \,   \; \frac{ \, 3}{  R } \; \dfrac{( 1 - c_s^2 ) \, c_s^2 }{(  3 \, c_s^2 - 1 )^2 } \; {\biggr\rvert }_{ {end}}  
\quad .
\label{rendxi2}
\eea
However the ratio $  R = \frac{ \, L}{  K^2 }  $ can be expressed in terms of the potential, using (\ref{potrrr}), 
from which using  the equations  (\ref{endofinf}), or  (\ref{whenend2}), we finally  get 
\bea
\rho_{end} \, = \,   \; \frac{ \, 9 \,  M^2  }{  4 } \, ( 1 - c_s^2 ) \; {\biggr\rvert }_{ {end}}  \quad ,
\label{rendfinx}
\eea
a very handy relation.  Note that all quantities are evaluated at end of inflation.  Using the lower bounds on $c_s^2$, see(\ref{boundc}), we get a strict  upper bound on $ \rho_{end}  $,
\bea
\rho_{end} \, < \,  \dfrac{ 3 \,  M^2 }{ 2 } \,   {\biggr\rvert }_{ {end}}   
\label{bbb1}  
\eea
In Eqs.    (\ref{rendfinx}) and    (\ref{bbb1}) the quantity $M^2$ is not constant, in general, 
but a function of the position, $h$ ( or $\phi$), which should be replaced by its  end of inflation value, as well. The actual upper bounds, which are extracted by solving the pertinent differential equations numerically,  may be smaller than those of Eq. (\ref{bbb1}). In fact $c_s^2$, at the end of inflation, depending on the model and  the values of the parameters involved, may be close to unity.  This is indeed the case when the effect of the $L$ - terms is small throughout the inflation evolution.  It may also happen that the lower bound in  (\ref{boundc}) may be larger due to the fact that $ L \dot{h}^2 / K  $ can never exceed some critical value. In that particular case the upper bound (\ref{bbb1}) is lowered.  This is the case, for instance,  in the minimally coupled models to be discussed below.

In general, there is a critical time, say $t_{crit}$, for which $\ddot{h} = 0$, that is the acceleration of $h$ vanishes. Note that this does not imply that the corresponding  acceleration for the field $\phi $ vanishes at $t_{crit}$
\footnote{
In fact when $\ddot{h} = 0$ we have, 
\bea
\ddot{\phi} =  \dfrac{1}{2 K } \,  \dfrac{ d K  }{  d \phi} \, \dot{\phi}^2 \neq  0 
\quad .
\eea
} .
%%%%%
From initial stage of inflation till $t_{crit}$ the acceleration, and also the velocity, of $h$ are negative,  $\ddot{h} \leq 0$,   $\dot{h} < 0$. Therefore  in this time interval a solution of (\ref{eomun}) exists provided, 
\bea
S_{crit} \equiv 
 U^\prime_{eff}(h) 
+\dfrac{1}{4} \, ( 2 K^\prime + 3 L^\prime \, \dot{h}^2 ) \,\dot{h}^2   \,  \geq 0
\label{ctitis}
\eea
This puts an upper bound on the velocity  $\dot{h} $ if $  L^\prime < 0 $ which is indeed the case in a variety of models.  This is exemplified below for a class of models that have attracted much  interest. 

\vspace*{5mm}

\noindent
{\bf{ Minimally coupled $h^n$ - models }}

\vspace*{3mm}

These models belong to the class {\bf{ M1}}, discussed previously, characterized by 
constants $g$ and $M^2$,  and a potential $V(h)$ which is a monomial, 
\bea
g = 1 \quad , \quad M^2 = 1/3 \alpha \quad , \quad V(h) = \dfrac{\lambda}{n} \, h^n
\quad ( n  = \text{ positive integer } ) 
\label{monmod}
\eea
Then the  functions $K, L$ are given by,
\bea
K(h) = ( 1+ c h^n )^{-1} \quad , \quad L = \alpha \, K 
\quad .
\label{KLdep}
\eea
with the constant $c$ being defined by
\bea
c = \dfrac{ 4 \lambda \alpha}{ n } \quad .
\label{defc}
\eea
%%%%
The case $n = 2$   belongs to the class of the cosmological attractors \cite{Carrasco:2015pla,Carrasco:2015rva}, which is clearly  seen if one uses the canonically normalized field $\phi$ of (\ref{canon}), see reference 
  \cite{Antoniadis:2018ywb}. 
Thus $L$ is linearly dependent on $K$ through the constant parameter $\alpha$. The potential $U_{eff}$ is given by
\bea
U_{eff}(h) = \dfrac{1}{4 \, \alpha} \,  \dfrac{ c h^n }{ 1 + c h^n} 
= \dfrac{1}{4 \, \alpha} \,  ( 1 - K )  \quad .
\label{pomod1}
\eea
In this model the terms $  S_{crit} $, defined in (\ref{ctitis}), receive the form
\bea
S_{crit} =
U^\prime_{eff}(h) \, (  1 - 2 \alpha \dot{h}^2  -  3 ( \alpha \dot{h}^2)^2 )   
\quad .
\label{bouti}
\eea
Due to the fact that the derivative of the potential stays  positive until $h$ vanishes for first time, at $t_0$,  positivity of  $  S_{crit} $, 
in the interval $  t \leq t_{crit}$, entails
\bea
\alpha \dot{h}^2  < \dfrac{1}{3} \quad .
\label{newb1}
\eea
which surely applies at $t_{crit}$. 
Since at $t_{crit}$, where $\ddot{h} = 0$,  the velocity  gets its minimum value, and it is negative until it vanishes for first time   at $t_1$,   we have, 
\bea
 \dot{h}(t)  > - \, \dfrac{1}{\sqrt{3 \alpha}} \quad, \text{or}  \quad \;  u(h)  > - \, \dfrac{1}{\sqrt{3 \alpha}} 
 \label{newb2}
\eea
from begin of inflation until the velocity vanishes for the first time. This region includes  the whole inflationary period and therefore these bounds apply to end of inflation !  Whether  the  lower bound above is reached in inflationary era  depends on  inputs. 
Using (\ref{whenend2}) and the bound (\ref{newb2}) one can derive
\bea
h_{end}^n \leq \dfrac{5}{3 \, c}  \quad  ,
\label{endhhh}
\eea
in this class of models. For large $\alpha$, corresponding to large $c$, when $\lambda$ is fixed,  the end of inflation value $h_{end}$, is small due to (\ref{endhhh}).
Note that the bound (\ref{newb2}) yields a more stringent lower bound on $c_s^2$ than the one  given by Eq. (\ref{boundc}). In fact one has, using (\ref{sound2}), 
\bea
c_s^2 > \dfrac{2}{3}
\label{twothird}
\eea
which on account of  (\ref{rendfinx}) results to 
\bea
\rho_{end} < \dfrac{3 M^2}{4} = \dfrac{1}{4 \alpha}   
\label{rhoend3}
\eea
which is actually half of the bound given by (\ref{bbb1}).  As we have already stated the actual upper bound is even less and depends on the precise value of $c_s^2$ at the end of inflation.  The right hand side of (\ref{rhoend3})  sets the maximum value the energy density $ \rho_{end}  $  can reach.  Obviously, this is almost saturated if $ c_s^2 \simeq  \dfrac{2}{3} $ at end of inflation, which using (\ref{omcs2}),  yields $ R \upsilon^2 \simeq 1/3 $, or $ \alpha u(h)^2 \simeq 1/3$. That is, at end of inflation the velocity of the $h$-field should be close to its lowest bound, as this is set by Eq (\ref{newb2}), for the energy density to reach its maximum allowed value in this type of models.  
%%% 
Then knowing the velocity at end of inflation, one can use (\ref{endofinf}), or same  
  (\ref{whenend2}), to derive  the value of $h_{end}$.   This  is found to  approach its upper bound  (\ref{endhhh}),    $  h_{end}^n \simeq \frac{5}{3 \, c} $. For large values of the parameter $\alpha$ this is indeed the case in this type of models. 
That is the bounds derived previously are saturated for sufficiently large values of $\alpha$. 
This will be exemplified in the following, when discussing the bounds set on the reheating temperature.

The aforementioned bound on $ \rho_{end} $ yields in turn bounds for the instantaneous reheating temperature 
$T_{ins}$  which is the largest reheating temperature allowed in any inflation model.  From Eq (\ref{tins}) we actually have,
 \bea
 T_{ins} =   \left( \dfrac{ 30}{\pi^2 \, } \dfrac{      \, \rho_{end} }{   \, g^{* (reh)} \,} \, \, \right)^{1/4}  \,  
 \leq \,  \left( \dfrac{ 15}{ 2 \pi^2 \,  \, g^{* (reh)}} \, \, \right)^{1/4} \, \alpha^{-1/4}
\label{tins2}
 \eea
 To convert it to $GeV$  this should be multiplied by the reduced Planck mass 
 $ m_P = (8 \pi G_N )^{-1/2} \simeq 2.435 \times 10^{18} \, GeV$.  With $ g^{* (reh)} =106.75 $ this yields the bound
 \bea
 T_{ins} \leq   
    \, \dfrac{ 0.290 \times m_P }{\,  \alpha^{1/4}} \, 
 = \,  \dfrac{ 0.7073 \, \times  10^{18}}{\,  \alpha^{1/4}} \, GeV
 \label{tins3}
 \eea

\begin{figure}[t]
\centering
%\begin{minipage}{.7\textwidth} 
  \centering     
   \includegraphics[width=0.49\linewidth]{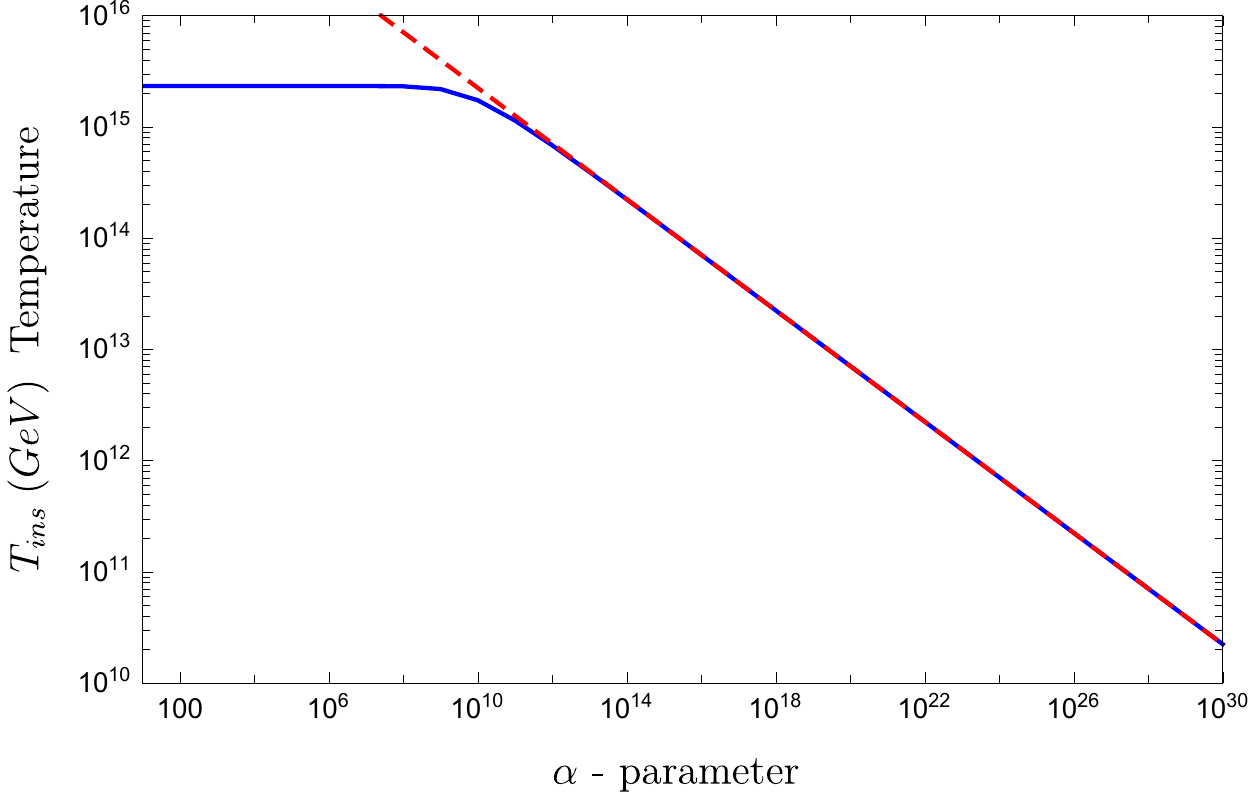}  
   \includegraphics[width=0.49\linewidth]{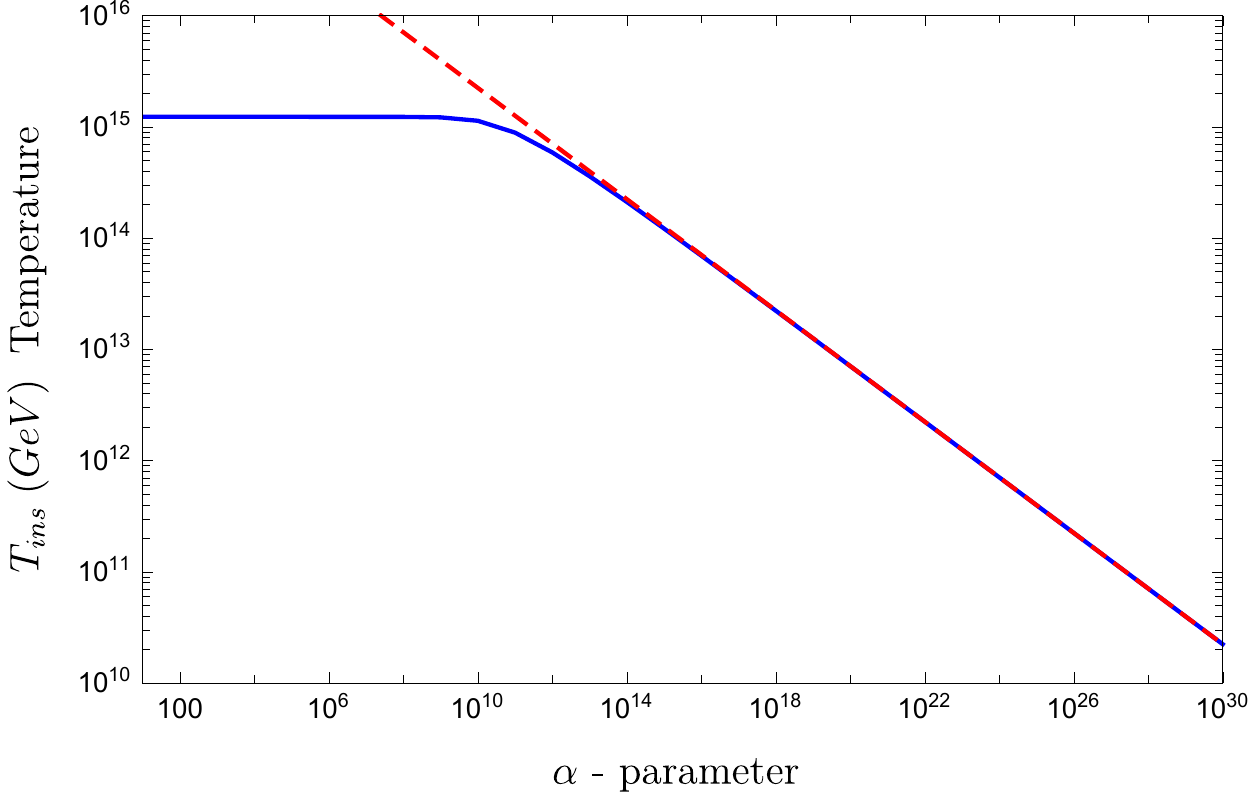}      
%\end{minipage}%
\caption{On the left,   we display the bound set on the instantaneous reheating temperature as function of $\alpha$ for the 
minimally coupled model with potential $ V(h) = {m^2} \, h^2 / 2$, and $ m = 6.2 \times 10^{-6}$.   On the right we show the same bound for the quartic potential $ V(h) = {\lambda} \, h^4 / 4
$, and $ \lambda = 2.025 \times 10^{-13}$. In both figures, 
the solid (blue) line is the actual bound as derived numerically and the dashed red line is the 
bound of Eq.  (\ref{tins3}). These coincide when $\alpha$ exceeds $\sim 10^{10}$.  
  }
   \label{figTins}
\end{figure}

In Figure   \ref{figTins} we display the actual  upper bound on $T_{ins}$ ( blue solid line ), derived numerically,  and  the strict mathematical upper bound of Eq. (\ref{tins3}), ( dashed red line ), which is based on (\ref{rhoend3}) .  
The cases displayed correspond to the minimally coupled model with $ n = 2 $ ( left pane) , with the parameter $\lambda $ 
of  the scalar potential in Eq. (\ref{monmod}) given by $\lambda \equiv m^2 $, with $ m = 6.2 \times 10^{-6}$, and the case $n=4$  ( right pane ), corresponding to  $ \lambda = 2.025 \times 10^{-13}$ . These values  are consistent with   scalar  perturbations, as we shall see.  

 Note that when $ \alpha  \lesssim 10^{10} $  the actual bound  is  lower than the mathematical upper bound set on $T_{ins}$ and   is  almost independent of $\alpha$,  depending however on the values of $m$ or $ \lambda$.  However,  the two bounds 
 coincide  for values  $\alpha > 10^{10}$.  
The reason is that in this region of $\alpha$ the sound of speed squared  $c_s^2$, at the end of inflation, approaches $2/3$, as we have already discussed, and the velocity approaches its lowest allowed limit, as this is  set by (\ref{newb2}), resulting to 
$ \rho_{end} \simeq 1 / 4 \alpha$. Note that in this case  the contribution of the $L$-terms is important in extracting the correct value of 
the field $h$ at the end of inflation, $h_{rend}$,  and hence the correct values for $ \rho_{end}$.  
In the regime of small $\alpha$,  smaller than $10^{10}$ or so, the sound of speed squared  $c_s^2$ at the end of inflation,  is very close to unity, that is the contribution of the $L$-terms is indeed negligible and the dynamics  is the same as in ordinary models, that is models which quartic in the velocity terms are absent.  It is only in this case that the usual slow-roll approximation schemes for extracting  $h_{rend}$ can be employed. 

%%%%%%%%%%%%%%%%%
The CMB observations restrict considerably  the predictions of all inflationary  models.  
The first calculations were performed in \cite{Starobinsky:1979ty,Mukhanov:1985rz,Mukhanov:1988jd} and  since then there has been an intense activity towards improving the calculations, by considering also  higher order corrections, demanded by the precise measurements  of the cosmological parameters, or   tackle theories with variable speed of sound, 
\cite{Lucchin:1984yf,Stewart:1993bc,Gong:2001he,Schwarz:2001vv,Martin:2002vn,Habib:2002yi,Leach:2002ar,Habib:2004kc,Casadio:2004ru,Wei:2004xx,Casadio:2005xv,Kinney:2007ag,Lorenz:2008je,Lorenz:2008et,Agarwal:2008ah,Martin:2013uma,Jimenez:2013xwa,Alinea:2015gpa}.
Therefore the bounds discussed previously on the instantaneous  reheating temperature are  narrowed if the observational constraints,  from various astrophysical sources are used. These actually impose bounds    on the parameter $\alpha$, or equivalently on the scale of inflation. For the derivation of the cosmological parameters, we shall make use of 
the number of e-folds $N_*$, corresponding to a pivot scale  $k^*$,  which can be  written as 
  \cite{Liddle:2003as,Dodelson:2003vq}, 
  \bea
 N_* \, = && \, 66.89 - ln c_s^* - ln\left( \dfrac{k^* }{  a_0 H_0} \right) + \dfrac{1}{4} \left( ln \, \dfrac{ \,  3 H_*^2}{ m_P^2} + ln \, \dfrac{ 3 H_*^2 m_P^2}{  \rho_{end}}  
\right)  - \dfrac{1}{12} \, ln \, g_s^{* (reh)}  \, 
\nonumber \\
&& + \dfrac{1 - 3 w }{ 12 (1 + w)} \, ln \left(  \dfrac{\rho_{reh}}{\rho_{end}} \right)
  \, .
 \label{nfold2}  
 \eea   
 %%%, 
 See also \cite{Martin:2010kz,Lozanov:2017hjm}.  The advantage of using this, among other, is that incorporates information 
 on the reheating temperature through  the last term,  the so-called {\it{reheating parameter}}.  This  does not contribute when reheating is instantaneous, in which case $  \rho_{reh} = \rho_{end} $. It does not contribute either when $ w = 1/3 $, a well-known result.
 We have included in Eq. (\ref{nfold2})  the dependence on the variable speed of sound and  the reduced Hubble parameter is taken equal to $ h = 0.676$, for details see  see  \cite{Gialamas:2019nly}. 
 Given a pivot scale $k^*$ the end of inflation affects the $N^*$ through   $\rho_{end}$. Note that in our approach we solve numerically the pertinent background  equations, to find the values of the cosmic scale factor and inflaton  velocity, as functions of the  field values, and we use  (\ref{endofinf}), or (\ref{whenend2}), to find where inflation ends. To this goal  slow- roll approximation has not been invoked. In the sequel  we solve (\ref{nfold2}) to find the pivot  values, 
 $\phi^*$,  corresponding to first horizon crossing of the given scale $k^*$, through which the cosmological parameters are determined. In the slow-roll approximation this is implemented by using $ N^* = \int_{\scriptsize{\phi_{end}}}^{ \tiny{\phi^*}}\frac{U \, d\phi}{ U^\prime }$, where $U$ is the inflaton potential.
 Concerning the cosmological observables, 
 for a  pivot scale,  $k^*$ that exited the sound horizon at $t^*$, that is $k^* c_s(t^*) = a(t^*) H(t^*)   $,  the scalar and tensor power spectra can be expanded about this pivot. Keeping  the first order terms in the Hubble flow functions (HFF), one has that 
the corresponding amplitudes are given by, 
\bea
A_s \, = \, \dfrac{H_*^2}{8 \pi^2 m_P^2 \, \epsilon_1^* c_s^*}  \,
 \left( 1 - 2 ( D+1) \, \epsilon_1^* - D \epsilon_2^* - (2+D) \, s_1^*
 \right)
 \, \,  , \,
 A_t \, = \, \dfrac{2 \, H_*^2}{ \pi^2 m_P^2 }  \,
\left( 1 - 2 ( D+1 - ln c_s^*) \, \epsilon_1^*  
 \right)  \, .
\label{power}
\eea
These can be found in \cite{Martin:2013uma}, where techniques similar to the WKB approximation has been used for their derivation. 
The constant $D$ is given by  $ D =7/19 - ln 3$. From these one has for the  tensor-to- scalar ratio,  
\bea
r \equiv \dfrac{A_t}{A_s} = 16 \,  \epsilon_1^* \,  c_s^* \, ( 1 + 2 \, ln c_s ^*\,  \epsilon_1^*  + D \epsilon_2^* 
+ (2 + D) \, s_1^*)   \quad .
\label{ratiorcor}
\eea
 Concerning the spectral index, following standard definitions, to  same approximation, we have
\bea
n_s  \, = \, 1  -2 \, \epsilon_1^* - \epsilon_2^* -  s_1^*  \, .
\label{speind}
\eea
 %%%%%%%%%%%%
 In the equations given before $ \epsilon_2, s_1 $, whenever they appear,  are given by 
 $  \epsilon_2 =   \dot{\epsilon}_1 / \epsilon_1 H$  and   $  s_1 =   \dot{c}_s / c_s H$.
 %%%%%% XXXXX
 For the  primordial power spectrum we  consider scales $k$ sampled by Planck CMB observations, lying in the range $ \sim 10^{-3} - 10^{-1} \, Mpc^{-1}  $.  The spectrum features small amplitudes   $ A_s \sim 10^{-9}  $ and is almost  almost scale invariant.
  \footnote{
 Starobinsky-type  models in metric formulation, with non-minimal coupled scalar field, can  yield enhanced primordial curvature perturbations on small scales \cite{Pi:2017gih}. Such a study  is however outside the scope of this article.
 } 
 
 %%%%%%

\begin{figure}[t]
\centering
%\begin{minipage}{.7\textwidth} 
  \centering    
 \includegraphics[width=0.49\linewidth]{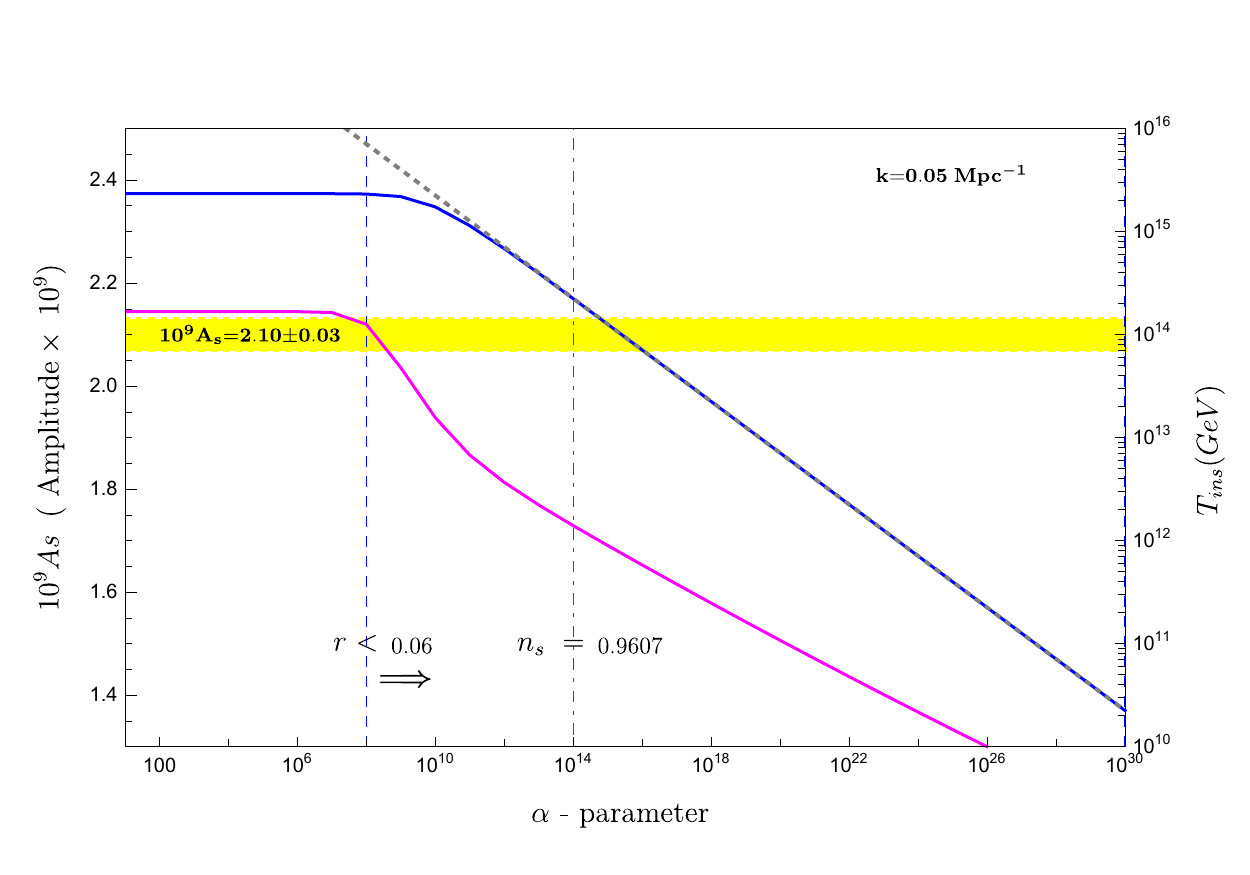}   
   \includegraphics[width=0.49\linewidth]{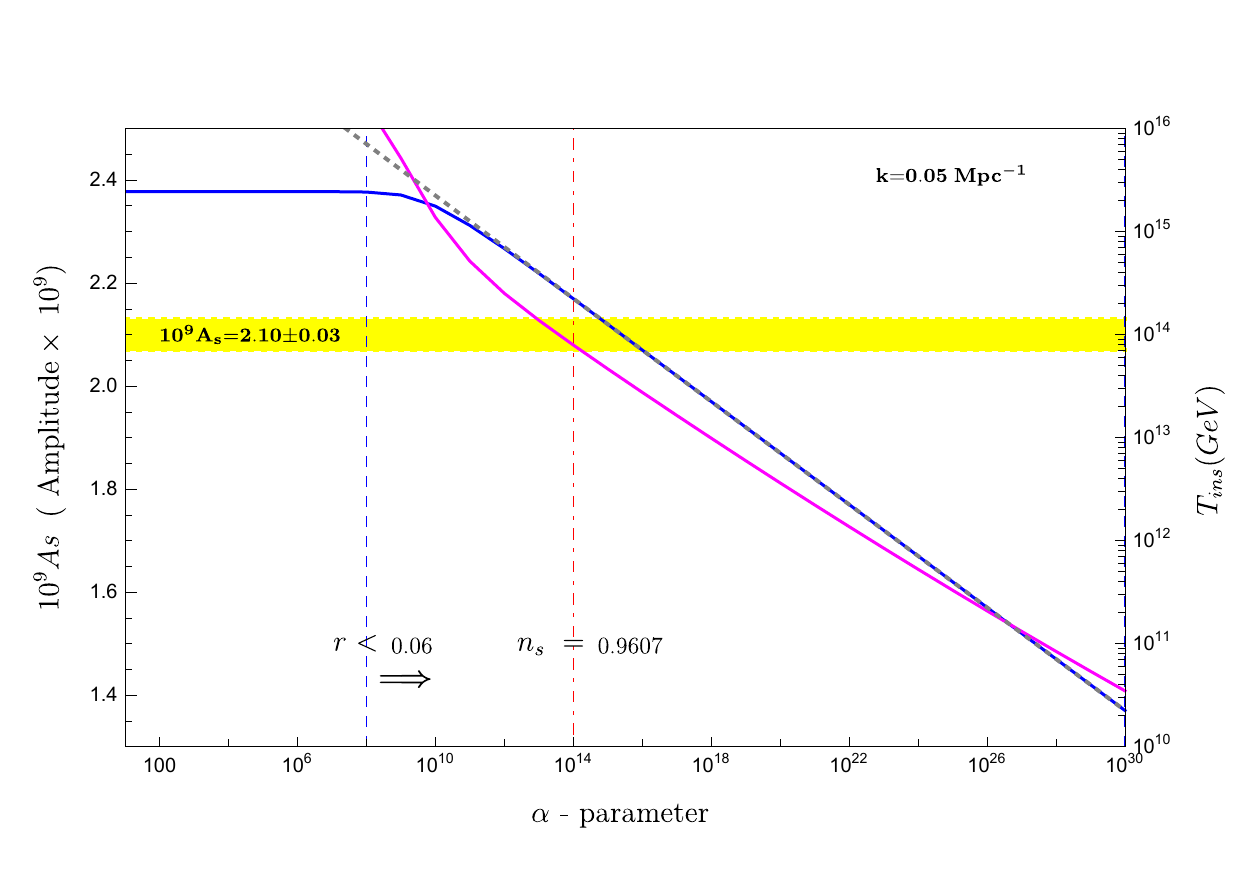}    
%\end{minipage}%
\caption{The bound set on the instantaneous reheating temperature, and other observables,  as function of $\alpha$ for the 
minimally coupled model with potential $ V(h) = {m^2} \, h^2 / 2$, for  $ m = 6.2 \times 10^{-6}$ (left) and 
$ m = 6.8 \times 10^{-6}$ (right). 
In these figures,   we display the amplitude of scalar perturbations $A_s$, in magenta, with values on left vertical axis, as function of the parameter $\alpha$, for  $  k^*$  shown in the figure, assuming instantaneous reheating,  $T_{ins}$, drawn in solid blue line, with values shown on the  right vertical axis. The horizontal stripe ( in yellow ) marks the region allowed by Planck 2018 observations on $A_s$. 
 Along we show the lower observational  bound on the  spectral index $n_s$, red dash-dotted vertical line line, and the bound  
 $r < 0.06$ set on the  tensor to scalar ratio, blue dashed line, in each of the  figures.
  }
   \label{figAs}
\end{figure}

In Figure   \ref{figAs}, we display the  instantaneous temperature $T_{ins}$ and the bounds set by other observables, assuming that reheating is instantaneous, that is   $T_{reh}=T_{ins}$. This is drawn in solid blue line with values  shown on the  right vertical axis. 
The gray dashed line is the bound on the instantaneous temperature discussed previously, see Figure \ref{figTins}.
These figures are the predictions for the minimally coupled model model 
  with potential $ V(h) = {m^2} \, h^2 / 2$, and slightly different values  $ m = 6.2 \times 10^{-6}$ (left pane) and  $ m = 6.8 \times 10^{-6}$ (right pane).   
In these figures,  the amplitude of scalar perturbations $A_s$  is displayed ( in magenta )  as function of the parameter $\alpha$, with values shown  on left vertical axis,, for a pivot scale $  k^* = 0.05\, Mpc^{-1}$.   The horizontal stripe, in yellow, marks the region allowed by Planck 2018 observations on $A_s = ( 2.10 \pm 0.03 \,) \times 10^{-9} $,  \cite{Planck:2018jri,Planck:2018vyg}.  
One notices that $A_s$ puts severe constraints on the parameter $\alpha$, and hence on the instantaneous temperature as well,  given the value of the parameter $m$. 

 Along we show the lower experimental  bound on the  spectral index $n_s$, as a red dash-dotted vertical line. The region to the left of this line is within the Planck 2018  limits $ n_s = 0.9649 \pm 0.0042 $. The bound $ r < 0.06   $  on the tensor to scalar ratio $r$,  set by various measurements, is also shown as a vertical blue-dashed line. The allowed region by the $r$-bound stands to the right of this line, designated by an arrow.  The aforementioned bound on $r$  is pretty close to that established 
 by Planck 2018 data, when combined with the BICEP2/Keck Array BK15 measurements, see  
 \cite{Planck:2018jri,Ade:2018gkx}.  The BICEP/Keck collaboration 
 \cite{BICEP:2021xfz} has further put  a more tightened  bound $ r < 0.036$. However we shall prefer to use the more conservative  bound $ r < 0.06   $.  Actually using the bound $ r < 0.036$ will shift very little  the blue-dashed line to the right and therefore the predictions are almost intact.

  The combined $n_s, r $ bounds further narrow the  range of  $\alpha$-values allowed  by  $A_s$.  We remark  that these bounds,  unlike those stemming from  $A_s$,  are rather insensitive to small changes of  the parameter  $m$ and for this reason the location of the $r$ and $n_s$ bound are almost same in  in the two figures. 
 However $A_s$ predictions depend sensitively on $m$. In the displayed figures the $A_s$-curves have a quite different shape,  although the parameter $m$ has only  slightly changed.  Notice that the case  $ m = 6.8 \times 10^{-6}$, corresponding to the figure on the right,  can be considered as the largest allowed by the  $A_s$ constraint,  if the bound on $n_s$ is observed,   assuming instantaneous reheating.  Actually for larger  $m$-values the allowed by $A_s$ region lies to the right of  the $n_s$ lower observational limit, shown by the dashed-dotted red line in the figures, and hence  excluded.  On the other hand the case $ m = 6.2 \times 10^{-6}$ is the minimum allowed, as lower values move the allowed $A_s$-region to the left of the   $ r < 0.06 $ bound.  
 Therefore combining all bounds we have that the allowed window for $m$ is in the range $m \simeq (6.2 - 6.8) \times  10^{-6}$ with corresponding limits for $\alpha$ in the range  $\alpha \simeq  10^8 - 10^{14} $.  These induce the bounds 
 $  T_{ins}   \simeq 2.2 \times 10^{14} GeV $ and $  T_{ins}   \simeq   2.3 \times 10^{15} GeV$ respectively, with the larger (smaller) bound corresponding to the smaller ( larger )  allowed value of $\alpha$. 
 We remark that for  $m$-values in the range given before, most of the  the allowed $\alpha$ range, actually  $\alpha > 10^{10}$, is within the regime where the quartic in the velocity terms are sizable and  $c_s^2 \simeq 2/3$ at end of inflation. 
  Note however that  the first horizon crossing of the CMB scales occurs when $c_s^2 \simeq 1 $, that is when the quartic terms are negligible. 
 Therefore, when 
 $\alpha > 10^{10}$, the previously discussed bounds  could not have been predicted in the  usual  slow-roll approximation schemes. In fact inflaton leaves the slow-roll regime well before the end of inflation, for such high values of $\alpha$.  As a result,  a more delicate study is needed in deriving the end of inflation energy density $\rho_{end}$ which determines the instantaneous reheating temperature, which we do in the present work. 
 It may not have been passed unnoticed that for $\alpha \lesssim 10^{9}$ the amplitude $A_s$, and also the temperature $T_{ins}$,  are insensitive to changes in $\alpha$. This is due to the fact that for such values of $\alpha$ the quartic in the velocity terms are small  and slow-roll is a good approximation scheme. Then using the slow-roll function $\epsilon_V$ to find the the pivot value $h^*$, or $\phi^*$,and end of inflation $h_{end}$, or $\phi_{end}^*$, yields results that are accurate enough. That done one finds that $A_s , T_{ins}$ depend explicitly on the parameter $m$ and only implicitly on $\alpha$.  That was shown analytically in  \cite{Gialamas:2019nly}. This is the reason for fixed $m$-values these quantities are almost constant with changes in $\alpha$. The same holds for other models, as well, namely the Higgs model, which we shall discuss in the next section.

Therefore,  in the framework of Palatini ${\cal{R}}^2$ inflation, in the context of the minimal models with  monomial potential $ V(h) = m^2 h^2 /2$, and $g=1, M^2 = 1 / 3 \alpha$,  the scalar power spectrum amplitude $A_s$ constrains the parameter $m$, which  in conjunction with $n_s, r \, $-data leads to  bounds on $\alpha$.  
Upper bounds on the instantaneous temperature can be established, which are saturated for large $\alpha$.  
For such values of $\alpha$ need go beyond slow-roll to derive reliable cosmological predictions.  
Note that by reinstating units,  the plateau values for the potential can be cast in the form $ U_{eff} (h) = 3 M_s^2 m_P^2 /4 $,   for a comparison with  Starobinsky inflation, where  for the case at hand  the  inflation scale $M_s$ is defined by 
 $ M_s = m_P / \sqrt{3 \,\alpha} $. Then the previously discussed  bounds on $\alpha$ translate to   $ M_s \simeq	  (  10^{-4} - 10^{-7}  ) \, m_P$.  For comparison, recall that CMB amplitude restricts the Starobinsky scale to $M_s = M_{Starobinsky} \simeq 1.3 \times 10^{-5} \, m_P$.
 
\begin{figure}
\centering 
%\begin{minipage}{.7\textwidth}   
  \centering  
    \includegraphics[width=0.49\linewidth]{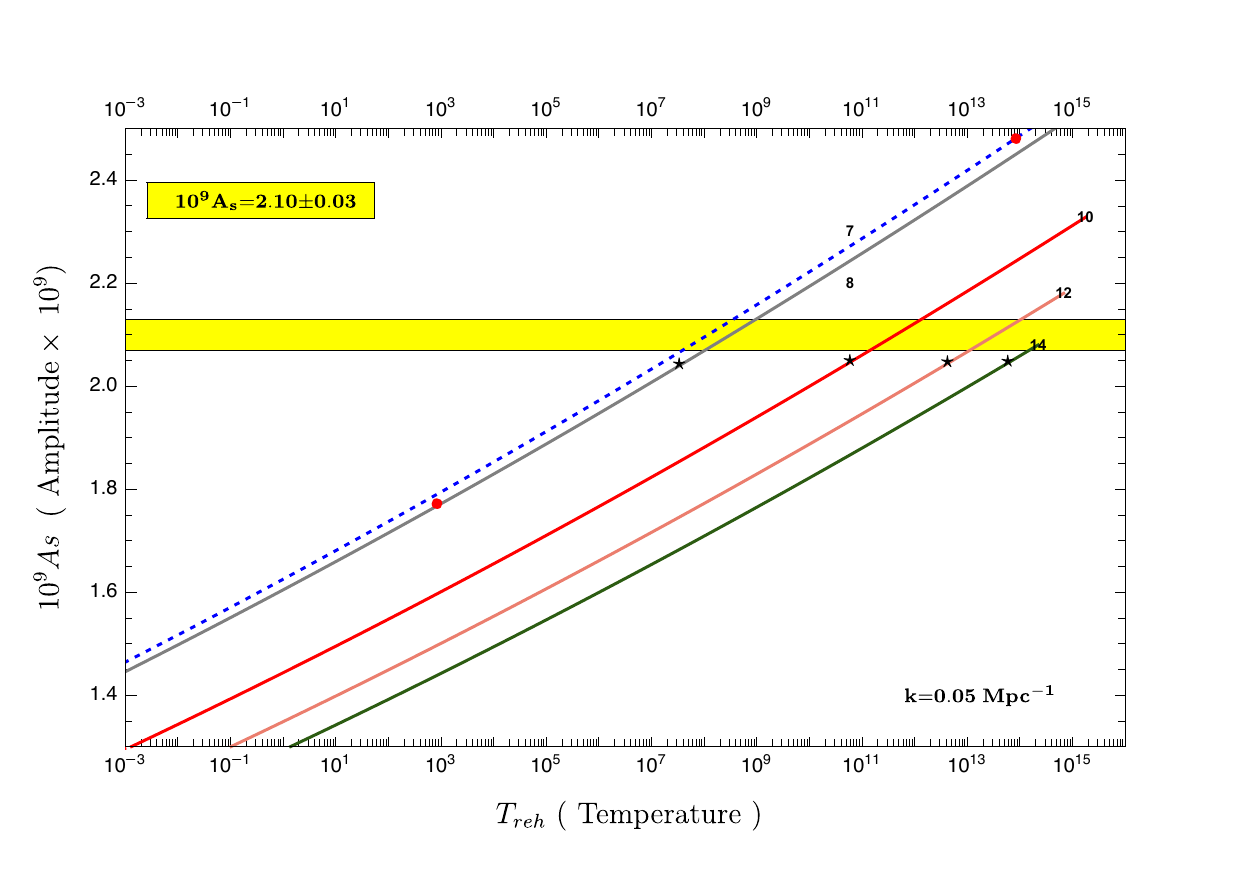}   
   \includegraphics[width=0.49\linewidth]{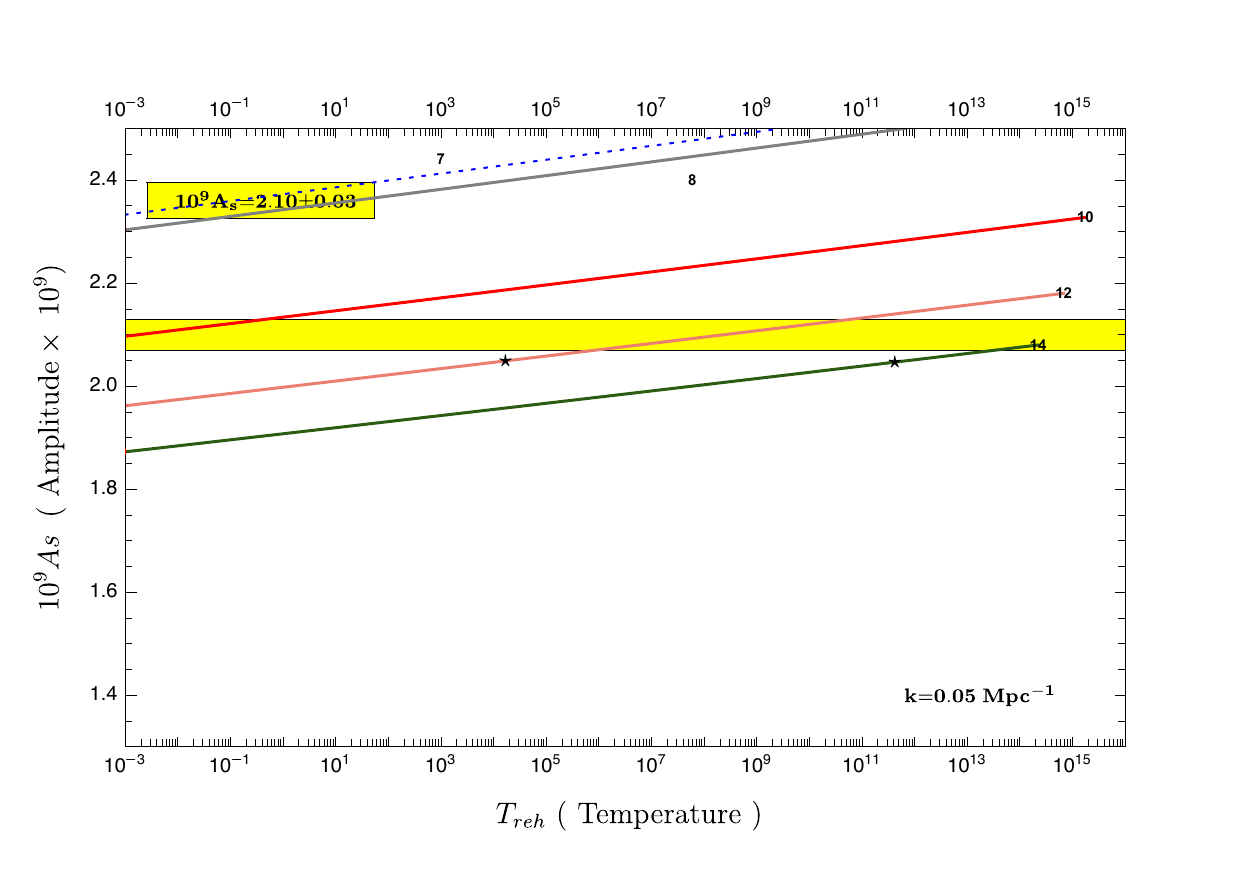} 
   \includegraphics[width=0.49\linewidth]{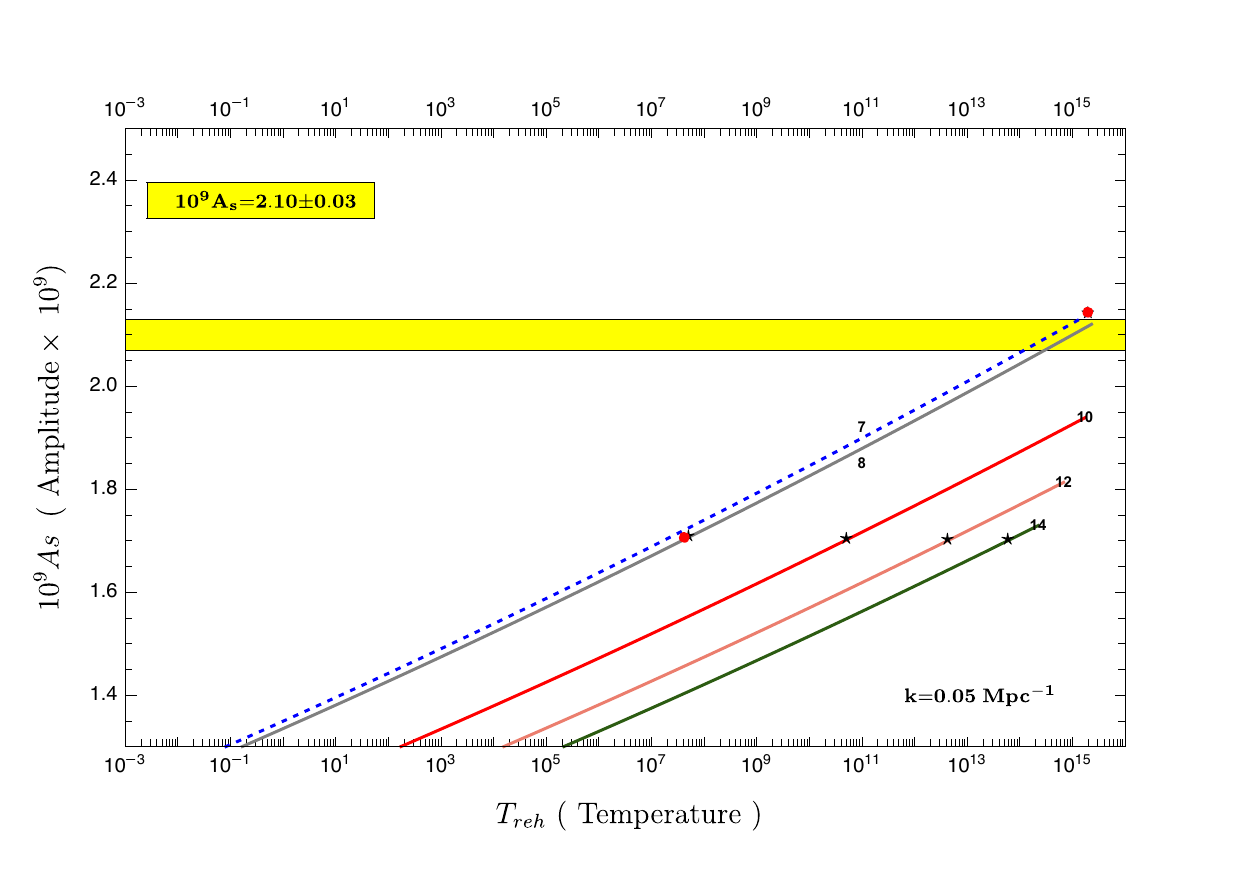}   
   \includegraphics[width=0.49\linewidth]{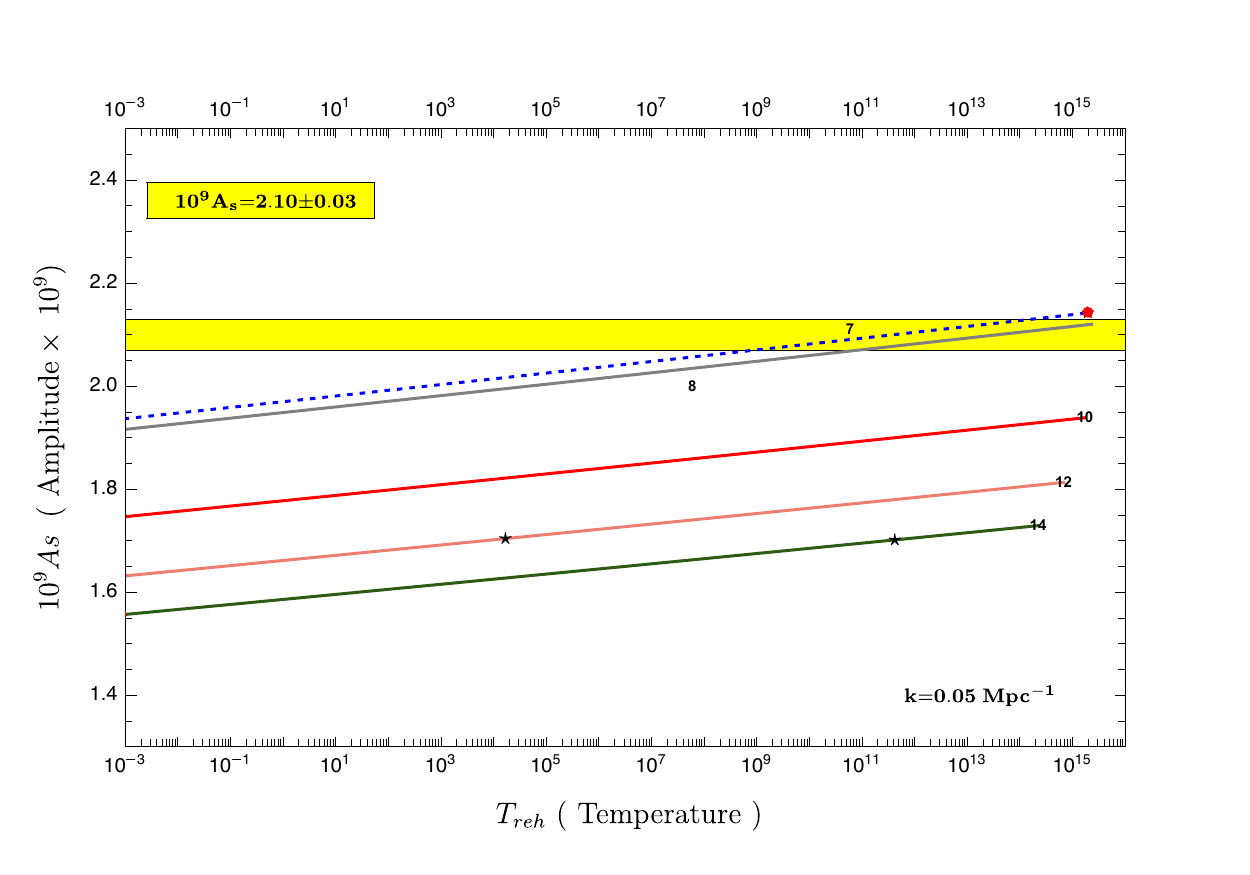} 
%\end{minipage}%
\caption{The amplitude $A_s$ as function of the reheat temperature, for various values of $\alpha$, 
marked by an integer on each line denoting the value of $ \log_{10} \alpha$, 
 for the minimally coupled model  with potential $ V = m^2 h^2 / 2 $.  
The equation of state parameter equals  to $ w =0.0$, left pane, and  $ w =0.25$, right pane.  At top, the parameter $m$ has its highest allowed value, $m = 6.8 \times 10^{-6} $ and at bottom $m$ gets its lowest allowed value, $m = 6.2 \times 10^{-6} $ 
The star symbol, whenever appears, indicates the point for which $n_s = 0.9607$, the lowest allowed observational bound. The red bullet, if it appears,  marks the boundary $ r = 0.06$. Points on the line lying  to the left of these symbols  are not allowed. 
   }
   \label{figTemp}
\end{figure}

 %% \newpage 
%%% Lower T %%% 
So far we have assumed that reheat is instantaneous and derived bounds arising from cosmological observations. 
By the same token,  similar bounds  can be also established for lower temperatures, as well.  The analysis, in this case, depends on the effective equation of state parameter $w$ . 
  We shall assume that the latter takes values in the range $ w \simeq 0.0 - 0.25$, with $w=0$ corresponding to the canonical reheating scenario. 
  This range of $w$ is favoured   in some inflationary scenarios, as we have already discussed. 
For the minimally coupled model, with potential $ V(h) = {m^2} \, h^2 / 2$,
  we have  found that  the bounds established for the parameter $m$, $ \,  6.2 \times 10^{-6} \lesssim m \lesssim  6.8 \times 10^{-6} $, hold for lower temperatures as well. Values outside this range are ruled out by the cosmological data. In fact lower $m$-values lead to $ 10^9 A_s \lesssim 2.0 $, while for  larger values,  either $A_s$ is  small or,  even when $A_s$ is within observational limits, $n_s$ is too low.  Hence $m$-values outside the   aforementioned range are discarded.  
  For any $m$, within this range, the amplitude $A_s$ and/or $r$,  the tensor to scalar ratio, put lower bounds on $\alpha$, while upper bounds on $\alpha$ are set by $A_s$ and/or $n_s$.    
  Given the parameter $m$, a range of allowed  $\alpha$ is therefore  established, and for any $\alpha$, in this range, the reheating temperature $T_{reh}$ takes values within an interval whose lower/upper limits correspond to the minimum/maximum allowed $A_s$. These limits are larger (smaller)  the larger (smaller) the value of $\alpha$ is, within its allowed range.   
  This is exemplified in Figure \ref{figTemp}, where  for the 
    largest and smallest allowed values  of $m$,   $m_{max} \simeq 6.8 \times 10^{-6}$ and  $  m_{min} \simeq 6.2 \times 10^{-6}   $,  we have drawn   
  the  amplitude $A_s$, as function of the reheat temperature, for different values of the parameter 
  $\alpha$.  Each line shown  is marked by an integer which denotes the value of $ \log_{10} \alpha$.
  The right end of each $A_s$ - line stops at the instantaneous  reheating temperature. 
  Whenever a star symbol appears on a line it is there to indicate the boundary  $n_s = 0.9607$. On the other hand whenever a red bullet appears it designates the location of the upper bound   $ r = 0.06 $.  
  Points on the line lying  to the left of these symbols  are not acceptable.   Whenever any of these  symbols is absent  there is no restriction stemming from the corresponding bound. 
  
  On the left panes of Figure \ref{figTemp},  the equation of state parameter is $w=0.0$ and on the right panes $w=0.25$.  
  For each $\alpha$, the segment of each line within the yellow stripe,  which designates the allowed range $A_s  =  2.10 \pm 0.07$, projected onto the $T_{reh}$ - axis locates the allowed range of temperatures for this value of $\alpha$.  For instance,  for 
the case shown on top left of this  figure, when  $\alpha = 10^{10}$ the temperature range is $ \simeq 10^{11} - 10^{12} \, GeV $. 
  As far as values of $\alpha$ are concerned, for the same case, 
  the allowed range of $\alpha$ is $\alpha \simeq 10^{8} - 10^{14}$. Lower values are excluded since they violate the $ r  < 0.06 $ bound.  As a demonstration of it,  we have displayed the $\alpha = 10^7$ case, by a blue dashed line, which while  being in agreement with 
  $n_s$ - data,  and also $A_s$ in some temperature range,   the bound on  $ r $  is violated for any temperature.  This is indicated by the red bullet on the far right of it. The same holds for lower  $\alpha$-values. Therefore  only values $\alpha \gtrsim10^8$ are allowed. 
  For the figure at the bottom right, the allowed range   is $\alpha \simeq 10^{10} - 10^{14}$.  Only values within this range can be compatible with $A_s$ observational limits, as is clearly seen. Note that,  although the $\alpha$ - range has shrunk, in comparison with the $w=0$ case,  it allows for lower temperatures. This will be discussed in the sequel.

 %%%%%%%%%%%%%%%%%%
 %%  \newpage
Concerning the highest temperature attainable, for any given value of the parameter $m$,  there is always a narrow range  of 
$\alpha$'s for which their corresponding instantaneous temperature are cosmologically  acceptable,  in the sense of giving predictions compatible with all data. See for instance Fig. \ref{figAs}. 
Since the instantaneous temperature $T_{ins}$ drops with increasing $\alpha$,  the  lowest of these 
$\alpha$'s,  yields the larger $T_{ins}$  for this particular value of $m$.  On the other hand, 
the minimum value  of the parameter $m$,    $m_{min} \simeq 6.2  \times 10^{-6}$, yields the lowest possible $\alpha$. Therefore,  
applying this reasoning for the minimum $m = m_{min}$, we pick the higher possible instantaneous temperature which is the highest possible reheating temperature. 
Phrased differently, and with reference Figure \ref{figTemp}, 
 for any given value of the parameter $m$ the higher temperature is the instantaneous  corresponding to the lowest $\alpha$ whose $A_s$ - line ends up within the yellow band $ A_s = 2.10 \pm 0.03$, shown in the figure, provided all other available data are observed.  Thus,  the highest possible reheat temperature  is the  instantaneous   corresponding to  the minimum value ,    $m = m_{min} \simeq 6.2  \times 10^{-6}$ and the lowest in this case $\alpha$,  having  its $A_s$ - line ending inside the yellow stripe, satisfying, also, the  $n_s, r $ data.  
    This is clearly shown at the  bottom panes of Fig. \ref{figTemp} from which  we see that the lowest $\alpha$ lies in a very narrow range around $\alpha \simeq 10^8$,  yielding  $T_{ins} \simeq 2.30 \times 10^{15} \, GeV$. This result is $w$ - independent, and for this reason the values of $T_{ins}$ are same for both left and right  figures presented. 
  Therefore the maximal attainable temperature within the context of this model is   $ 2.30 \times 10^{15} \, GeV$, quoted before, as being the largest instantaneous temperature, of all, for which all data are satisfied.  In order to complete the picture, for the maximum value 
  $m_{max} \simeq 6.8  \times 10^{-6}$ the situation is displayed at the  top of Fig. \ref{figTemp}. In this case the lowest allowed 
  $\alpha$, with the desired properties, is  $\alpha \simeq 10^{14}$, or somehow lower than it. This yields as instantaneous temperature $ 2.23 \times 10^{14} \, GeV$, obviously lower than the one corresponding to $m_{min}$ by an order of magnitude or so. 

%% \newpage

%%%%%%%%%
  As far as the lowest temperatures are concerned. It should not have passes unnoticed that when $w=0$, the available temperatures have lower bounds. From the left panes in Fig. \ref{figTemp}, at both top and bottom, this is clearly seen, and this is the case for any allowed value of $m$ in the canonical reheating scenario, i.e. when the equation of state parameter is vanishing. In the $w=0$ case the lowest temperature is reached when $m$ gets its maximum value. From the top left figure in Fig. \ref{figTemp}, we see that this cannot be lower than $ T_{reh} \simeq 10^8 \, GeV$,  obtained when $\alpha \simeq 10^8$, the minimum allowed.  
  For the minimum $m$, the lowest temperature is higher,  $ T_{reh} \simeq 10^{14} \, GeV$, as seen at the left bottom figure. Thus in this model and for canonical reheat, lower bounds on $T_{reh}$ are imposed,  and   temperatures below $ T_{reh} \simeq 10^8 \, GeV$  are  unattainable. However the situation drastically  changes if $w$ is allowed to take non-vanishing values. 
%%%
For $w=0.25$ and for the maximum $m$, at the top right panel of figure,  the lowest  acceptable $\alpha$ is  $\simeq 10^{10}$, and  temperatures in the range $ \simeq 0.01 \, MeV  - 0.6  \, GeV $ are allowed
\footnote{
  As a general  remark, concerning $w$,   the closer the $w$ is to the radiation value $1/3$ the more extended is the range of the allowed temperatures. Evidently for $ w = 1/3$, there are no restrictions imposed on the reheat temperature. }.
  %%%
Recall however that, low temperatures  are  constrained by BBN, which  imposes lower bounds $\simeq 1 MeV$ on the  reheating temperature.  For the minimum $m$ - case , see right pane at bottom, lower temperatures, as compared to $w=0$, can be obtained but not lower than$\simeq 10^{11} \, GeV$.

%%  \newpage
  %%%%%%%%%%%%%%
 %%%%%%%%%%%%%% MINI        
%%%%%% MINI 

 %%%END  Lower T %%%   
% \newpage

%%%%%%%%%%%%%% MINI        
%%%%%% MINI 

   %%    
%%%%%%%%%%

%%%%%%%%%%%%%% MINI        
\begin{figure}[H]
\centering
%\begin{minipage}{.7\textwidth} 
  \centering     
 \includegraphics[width=0.49\linewidth]{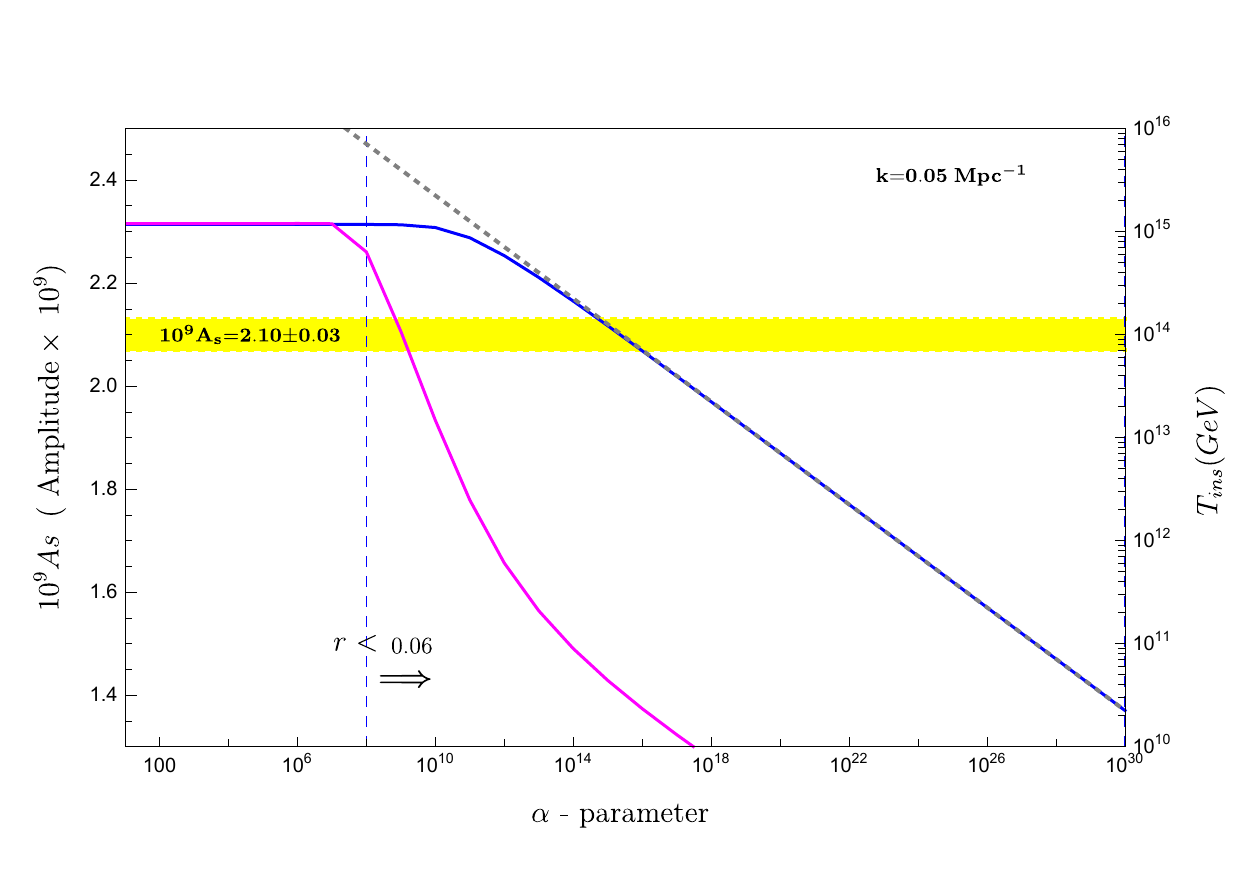}   
   %\end{minipage}%
\caption{The bound set on the instantaneous reheating temperature, and other observables,  as function of $\alpha$ for the 
minimally coupled model with potential $ V(h) = \lambda \, h^4 / 2$,  for 
$ {\lambda} \equiv m^2 $ with  $m=4.1 \times 10^{-7}$.
 Along we show  the bound  $r < 0.06$ set on the  tensor to scalar ratio, as a blue dashed line. In the whole range for $\alpha$ 
 the spectral index is below  observational limits.
  }
   \label{figAsQ}
\end{figure}  
%%%%%% MINI 

For the minimally coupled model with quartic potential  $ V(h) = {\lambda}  \, h^4 / 2$ the situation is similar as far as $A_s$ and $r$ constraints are concerned. However there is no agreement with $n_s$ data. As an example we display a representative case,
in Figure \ref{figAsQ}, for  $ {\lambda} \equiv m^2 $ with   $m=4.1 \times 10^{-7}$. As in the $ V(h) = {m^2} \, h^2 / 2$ case there is a lower bound on $\alpha > 10^8$, imposed by $r < 0.06$, and for the case displayed there is a also complete agreement with the amplitude $A_s$ for $\alpha $ slightly below $10^{10}$. However the spectral index  for any $\alpha$, allowed by  $A_s$,  is $n_s < 0.9490$, well below the observational limits.  Therefore the quartic potential fails to satisfy all observational data. The situation changes if the model is promoted to include non-minimal couplings, as is the case of the Higgs model to be studied in the next section.

 \vspace*{7mm}

\noindent
{\bf{ Non-minimally coupled models }}

\vspace*{3mm}

A particularly interesting model, belonging to the class {\bf{ M2 }}, is the one for which

\bea
g = 1 + \xi h^2 \quad , \quad M^2 = 1/3 \alpha \quad , \quad V(h) = \dfrac{\lambda}{4} \, h^4
\quad  
\label{monmod2}
\eea
 
This model is well-known to arise from the Higgs coupling to Palatini gravity in the unitary gauge when the electroweak scale is considered small. The parameter $\xi$ sets the coupling of the Higgs field to the curvature terms and $\alpha$ is the coefficient of the 
quadratic in the curvature term as defined in  (\ref{fr}).

%%%%%%%%%%%%% 

The Higgs coupling to gravity and its role as the inflaton,  in the metric formulation,  has been proposed in 
\cite{Bezrukov:2007ep,Bezrukov:2008ej}  and it has been  widely studied,  since then, 
 \cite{ Barbon:2009ya,Barvinsky:2009fy,Barvinsky:2009ii,Germani:2010gm,Germani:2010ux,Lerner:2010mq, Karam:2018mft,Jinno:2019und,Bezrukov:2010jz,Kamada:2010qe,Kamada:2012se,Bezrukov:2013fka, Allison:2013uaa,Bezrukov:2014bra,Hamada:2014iga,Hamada:2014wna, Salvio:2015kka,Saltas:2015vsc,Calmet:2016fsr,Jinno:2017lun, Bezrukov:2017dyv,Antoniadis:2018ywb,Antoniadis:2018yfq,Enckell:2018kkc,
He:2018gyf,Gundhi:2018wyz,Rubio:2018ogq,Karam:2021sno,Karam:2021wzz,Tenkanen:2020cvw,Lykkas:2021vax,Gialamas:2019nly,Gialamas:2021enw,Dioguardi:2021fmr,He:2018mgb,Steinwachs:2019hdr,
 Rubio:2019ypq,Takahashi:2018brt,Tenkanen:2019jiq,Cheong:2021vdb, 
 Hamada:2020kuy,Rigouzzo:2022yan,Durrer:2022emo,Dux:2022kuk,He:2022xef}  
 both in the context of the metric and Palatini  formulation. The importance of the $  {\cal{R}}^2 $ coupling, in the Palatini approach,  has been discussed in   \cite{Antoniadis:2018ywb,Antoniadis:2018yfq,Takahashi:2018brt,Tenkanen:2019jiq,Tenkanen:2020cvw,Lykkas:2021vax,Karam:2021sno,Gialamas:2019nly,Gialamas:2021enw,Dioguardi:2021fmr,Cheong:2021vdb}. 
 Quintessential inflation  in the framework of Palatini $  {\cal{R}}^2 $  gravity has been considered in 
 \cite{Dimopoulos:2020pas,Dimopoulos:2022tvn,Dimopoulos:2022rdp} 
 
  In this work we shall show that the quartic coupling $ \lambda$, as in the minimally coupled quartic model  studied previously, is constrained considerably by cosmological data, especially by the power spectrum amplitude $A_s$.  Combining all data,  
 further limits are imposed restricting the available options.   
%%%%%%%%%%%%%  
The pertinent functions $K, L$ are given by,    
\bea
K(h) = \dfrac{1+ \xi h^2}{ (1+ \xi h^2)^2+ c h^4 }   
\quad , \quad L = \dfrac{\alpha}{ (1+ \xi h^2)^2+ c h^4 } 
\quad .   
\label{KLdep2}
\eea
while the  potential $U_{eff}$ is given by, 
\bea
U_{eff}(h) = \dfrac{1}{4 \, \alpha} \,  \dfrac{ c h^4}{  (1+ \xi h^2)^2+ c h^4 } . 
\label{pomod2}
\eea
%%%%%%%%%%   

The bounds set on the instantaneous temperature are more difficult to derive in this case due to the dependence on the parameter 
$\xi$. Although the analysis is the same,  in this case the term in Eq.  (\ref{ctitis}) is not as simple as that given by (\ref{bouti}). For the case at hand we had better used $\omega = L \dot{h}^2 / K $ as dependent variable n Eq (\ref{eomunX}), instead of $u$. 
Thus is the same quantity used in (\ref{omcs2}). 
That done 
the pertinent equation takes on  the form, 
%%%%%
\bea
 ( 1 + 3 \, \omega ) \,  \dfrac{d\omega}{dh} - 3 H  \sqrt { \frac{4 L}{K} } (  1 +  \omega  ) \, \omega^{1/2} + 
   {\tilde{S}}_{crit}
 =0  \, ,
\label{eomunX3} 
  \eea
 where
 \bea
 {\tilde{S}}_{crit} = \frac{2 L}{K^2} \, U^\prime_{eff}(h) \left( 1 - 2 \omega -3 \omega^2 \right)  \quad .
 \label{crit3} 
 \eea
 %%%
 Modulo the factor $2 L / K^2$, this is reminiscent of (\ref{bouti}). 
 In  (\ref{eomunX3})  we have anticipated the fact that in the region of interest, from begin of inflation until the velocity $u$  vanishes for the first time,  $u$ is negative, and hence the negative sign in the Hubble term. Note that $\omega$ starts increasing, as $h$ decreases, while it vanishes  when the velocity $u$ vanishes for the first time, and therefore a maximum of $\omega_{max} $  is developed  at  a critical value $h_{crit}$. At this point   $ d \omega/dh = 0$, while   $d \omega/dh  \leq 0  $ for any  $ h \geq h_{crit}$, since $\omega$ increases with decreasing $h$, in this region. Therefore  from (\ref{eomunX3}) we deduce that $  {\tilde{S}}_{crit} > 0   $ for any  $ h \geq h_{crit}$, which   entails $\omega  < 1/3 $, in this region,  and therefore the maximum value of $\omega$ is bounded, 
 $\omega_{max}  < 1/3 $. Due to the fact that  $\omega_{max}$ is the maximum value we have
 \bea
  \omega  < 1/3 
  \label{omegamax}
 \eea
 %%%%
 for any $h$ from  begin of inflation until $u$ it vanishes.  
 This region  certainly includes the end of inflation $h_{end}$ and this puts an upper bound on the value of 
 $ \omega_{end} \equiv \omega(h_{end})$ at end of inflation, i.e. 
 \bea
 \omega_{end}  < 1/3 \quad .
 \label{omegaend}
 \eea
Note that this is the analog of (\ref{newb1}) for the minimally coupled models studied earlier. 
The bound (\ref{omegamax}) yields again the bound  of (\ref{twothird}), $c_s^2 > 2/3$,  which results to (\ref{rhoend3}), $ \rho_{end} < 1/4 \alpha $ . Therefore the bound    (\ref{tins3})  on the instantaneous temperature holds in the Higgs case as well. 

\begin{figure}[t]
\centering
%\begin{minipage}{.7\textwidth} 
  \centering     
 \includegraphics[width=0.49\linewidth]{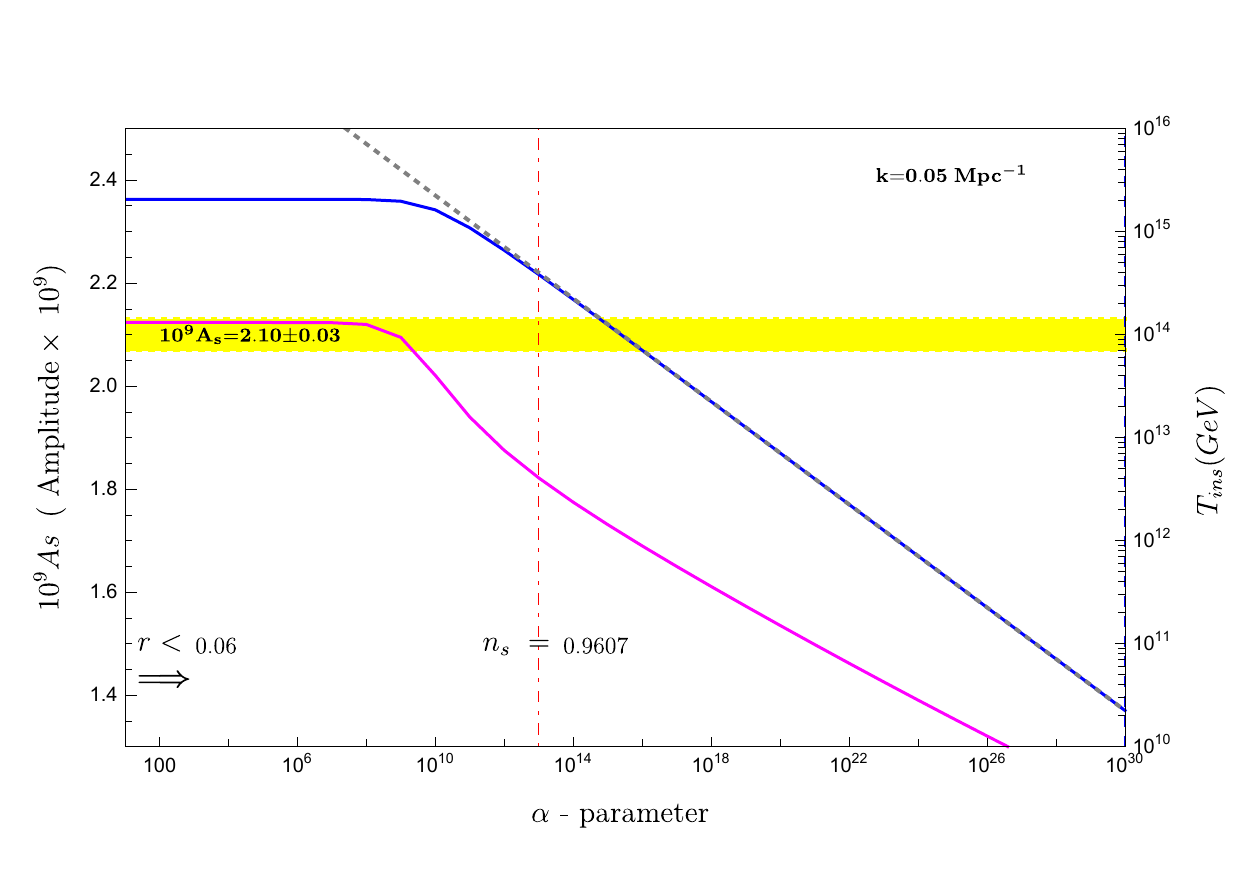}   
   \includegraphics[width=0.49\linewidth]{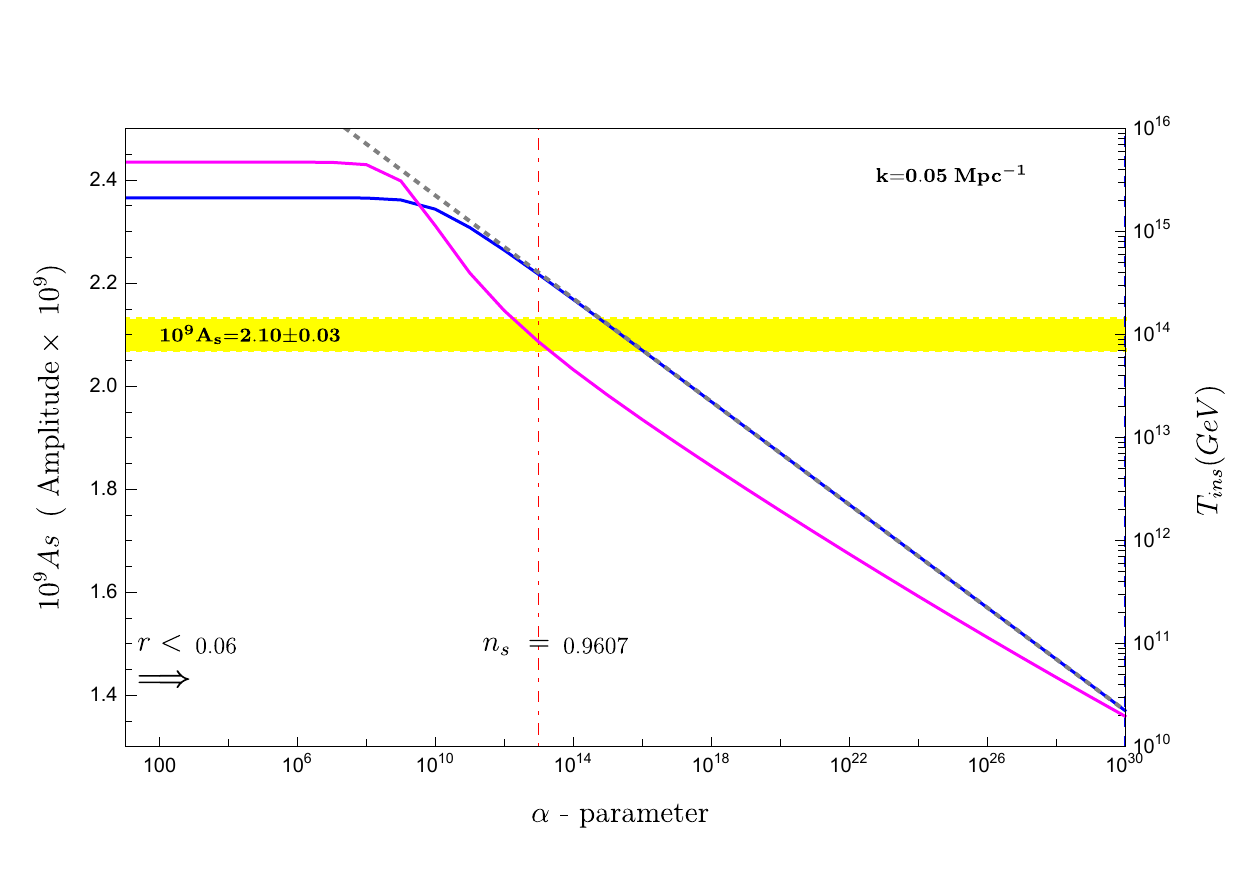}    
   %%%
    \includegraphics[width=0.49\linewidth]{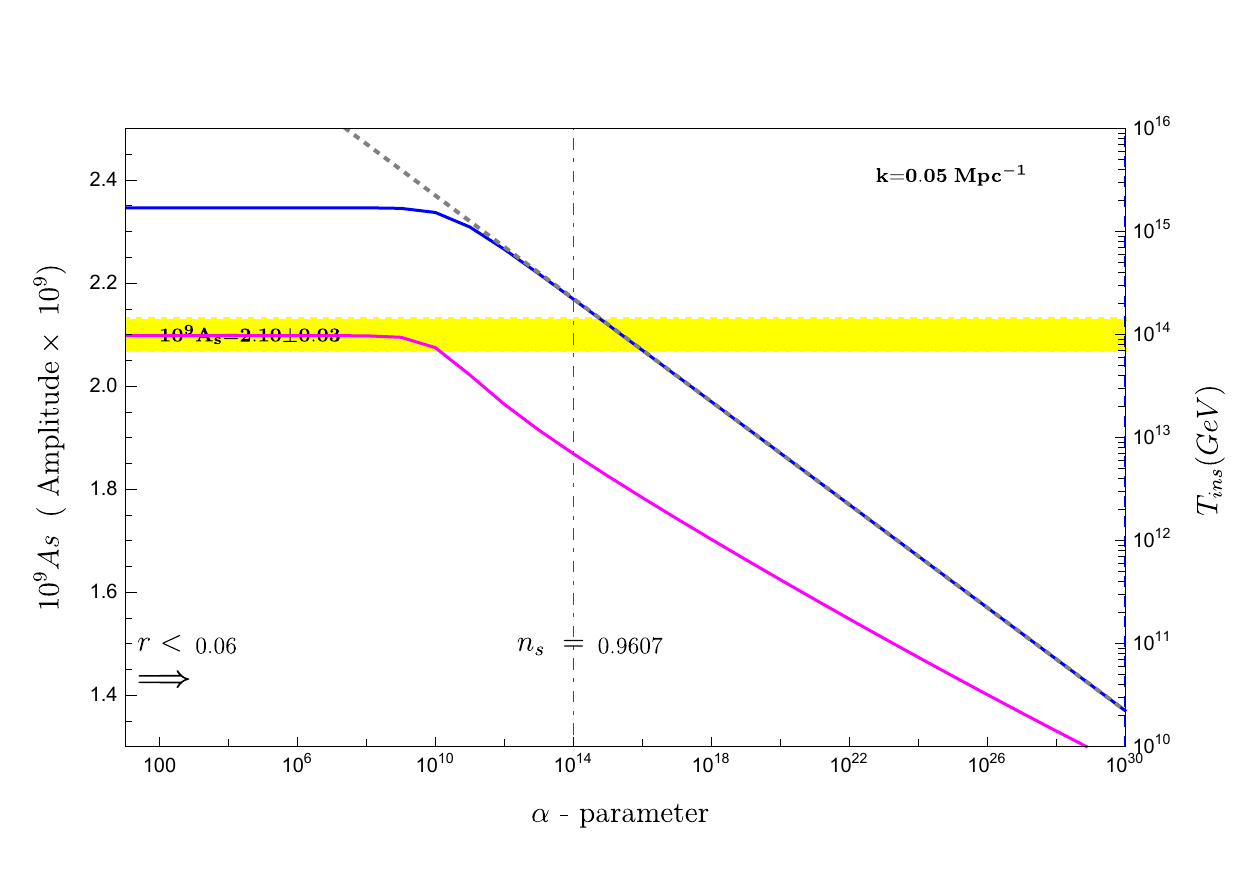}   
   \includegraphics[width=0.49\linewidth]{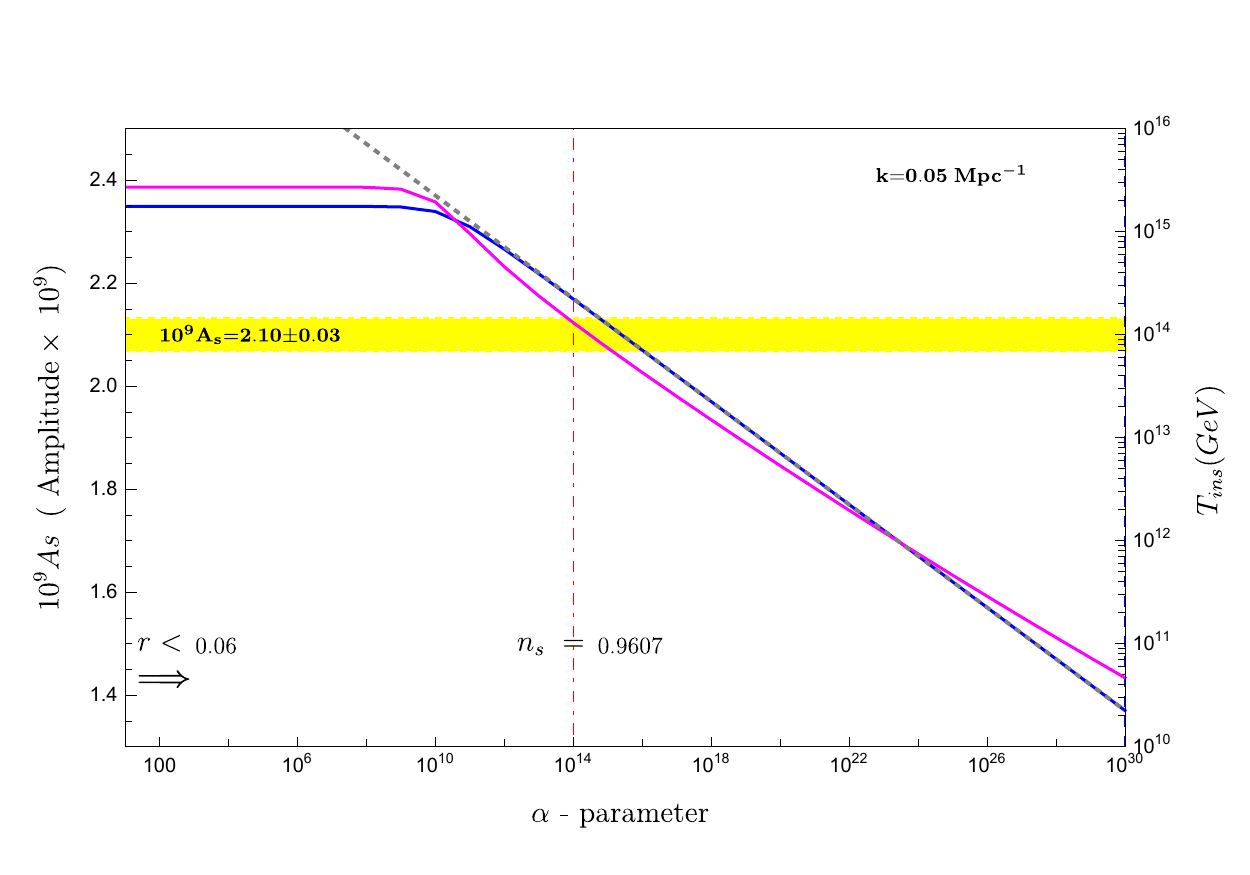}    
%\end{minipage}%
\caption{As in  Figure \ref{figAs},  we display 
the bound set on the instantaneous reheating temperature, and other observables,   for the Higgs model,  
$ V(h) = {\lambda} \, h^4 / 4$. We have set $\lambda = m^2$ and 
the figures on top correspond to  $\xi=0.1$.  For the  top left pane $m=2.85 \times 10^{-6}$ and for the right $m=3.05 \times 10^{-6}$.
These are the lowest and highest allowed  $m$-values  when $\xi=0.1$.
The figures  at bottom correspond to  $\xi=1.0$.  For the  bottom left pane $m=9.10 \times 10^{-6}$ and for the right $m=9.70 \times 10^{-6}$. These are the lowest and highest allowed  when $\xi=1.0$.
 Note that in all cases displayed  the bound  for $r $ has been moved to the far left so that all $\alpha$-region .
  is consistent with the bound $r < 0.06$ set on the  tensor to scalar ratio. 
  }
   \label{figAsHiggs}  
\end{figure}

\begin{figure}[H]
\centering
%\begin{minipage}{.7\textwidth} 
  \centering   
 \includegraphics[width=0.49\linewidth]{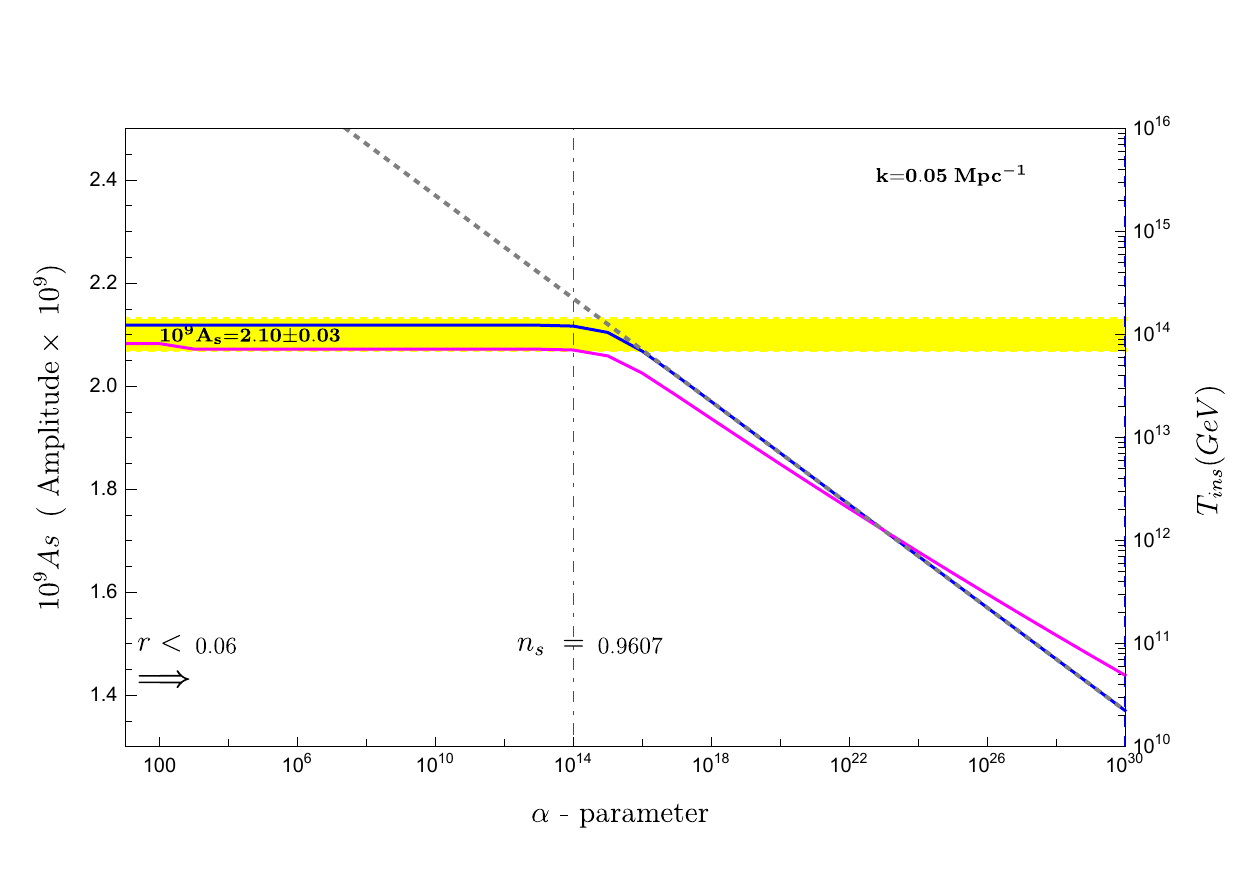}     
\caption{ As in in Figure \ref{figAsHiggs},  we display 
the bound set on the instantaneous reheating temperature, and other observables,   for the Higgs model,  
for a  large value $\xi=10^5$.  The case displayed corresponds to  $m=3.05 \times 10^{-3}$ which is the lowest allowed by all data for this value of $\xi$. A slight increase of  $m$ will move the $A_s$ predictions off experimental bounds.  Note that in this case the allowed region belongs to the  slow-roll regime.
  }
   \label{figAsHiggs2}  
\end{figure}

For sufficiently  large $\alpha$, as in the case $\xi=0$, the bound  
$\omega_{end} \simeq 1/3$ is saturated.  Then $ \rho_{end} = 1/4 \alpha $ and the upper bound on $T_{ins}$ of Eq.  (\ref{tins3}) is reached. 
With the aid of end of inflation relation  (\ref{whenend2}) we get in a straightforward manner 
$  R U_{eff} = 5/12$, at end of inflation, or using the form of the potential  as given in Eq.   (\ref{potrrr}),  $R_{end} = 8 \alpha / 3$.
This relation can be solved for $h_{end}$,  to derive the value of $h_{end}$, when  the upper bound on $T_{ins}$ is reached. The solution is,
\bea
( \sqrt{3 c/5} - \xi ) \, h_{end}^2 =1 \; .
\label{hfinis}
\eea
Note that this holds for sufficiently large $\alpha$.  We have verified that  Eq (\ref{hfinis}) indeed reproduces very accurately the numerical results for $h_{end}$,  for values $\alpha > \alpha_{crit}$,  with $\alpha_{crit}$   given by where the coefficient of $h_{end}^2$ on the left hand side of (\ref{hfinis}) vanishes. This results to a critical value
$  \alpha_{crit} = \frac{5}{3} \, { \left( \frac{\xi^2}{\lambda} \right) }$.  
A last comment concerns the value of $ \xi \, h_{end}^2 $ which turns out to be much smaller than unity in the regime $\alpha \gg \alpha_{crit}$.  In this case $g(h_{end}) \simeq 1 $ and hence  predictions are expected to be  same as with the $\xi=0$ case.
%%%%%%
Anticipating the fact that the Higgs model can be in agreement with cosmological observations, as we shall discuss in the sequel, this by no means should lead us to the wrong conclusion that the simple quartic potential, $\xi =0$, seen as the limiting case of the Higgs model, when  $ \xi \, h_{end}^2 \ll 1$,  can lead to successful inflation, In fact $ \xi \, h_{end}^2 = {\left(  \sqrt{ \alpha / \alpha_{crit}} - 1  \right)}^{-1}$, and one needs $ \alpha \geq 10^4 \, \alpha_{crit}$, or so,   to obtain indeed small values $  \xi \, h_{end}^2 \leq 10^{-2} $.
Such values for $\alpha$ outstrip the lower observational bound on $ n_s $, as shown in  Figures \ref{figAsHiggs},\ref{figAsHiggs2}, in which some representative outputs are displayed,  and therefore are not acceptable. This is in perfect consistency with the statement made towards the end of previous sub-section that the simple quartic potential, $\xi = 0 $ case, is in tension with $n_s$, predicting too low values for the spectral index.
%%%%%%%

Predictions  of the Higgs  Model  when the parameter $\xi$ is small,  $\xi=0.1$,  are shown at top  of Figure \ref{figAsHiggs}. Denoting 
the quartic coupling by $ \lambda = m^2 $,  the cases shown correspond to 
 the lowest, $ m = 2.85  \times 10^{-6}  $ ,  and largest, $ m = 3.05  \times 10^{-6}  $  allowed, when all observational data are observed and reheating is instantaneous. 
At the bottom of the same figure the case $\xi = 1.0$ is displayed. In this case the minimum and maximum allowed $m$- values are 
$ m = 9.10  \times 10^{-6}  $ and $ m = 9.70  \times 10^{-6}  $ respectively.  

Given $\xi$,  the  min/max values of $m$ and  the   ranges for the parameter  $\alpha$   are shown  in Table \ref{table111}. 
The corresponding ranges for the instantaneous temperature and  the cosmological observables
$ A_s, n_s, r,$ and the number of efolds $N_* $,  are also shown corresponding to a pivot scale $ k^* = 0.05 \, Mpc^{-1}$. 
In all cases the value of $r$ is small enough to  satisfy  the more stringent  bound $ r < 0.036$  put by BICEP/Keck  observations \cite{BICEP:2021xfz}. 
For the maximum allowed  value of $m$, the spectral index  $n_s$ is close to its lowest value, allowed by cosmological observations, while 
$\alpha$ gets its largest value. At the same time,  the  instantaneous reheating temperature gets its lowest value. 
These $\alpha$-values  are in the regime where the quartic in the velocity terms contribute substantially and   slow-roll approximation  is not applicable. 
Thus need go beyond slow-roll to derive these predictions, which has been done numerically.
The lowest $m$-values allow for a broad range   $\alpha \lesssim 10^9 $ and instantaneous temperature is larger. 
In this case,  the inflationary dynamics can be  successfully described by the slow-roll mechanism. The constancy of $A_s$ for the lowest allowed  $m$-values  actually follows from slow-roll. Using slow-roll approximation $A_s$ is found to depend on the ratio $c/ \alpha$, that is on the parameter $m$ alone  and not on $\alpha$. 
Note that for the  lowest $m$ case the tensor to scalar ratio is of the order of $ r \sim 10^{-3}$, much larger, by almost three orders of magnitude, from the largest $m$-case. These values will be  therefore within reach by future missions aiming to probe small values of $r$ as small as $ r \sim 10^{-3}$, or so, \cite{Kogut:2011xw,Matsumura:2016sri}.

One observes that the instantaneous temperature indeed touches the bound  (\ref{tins3}) for sufficiently large values of $\alpha $,    larger than some critical value $\alpha_{crit}$ that indeed coincides with the one estimated previously, after Eq.  (\ref{hfinis}), for each of the cases considered. Actually, 
$\alpha_{crit}$ separates the two regimes. The slow-roll and the region of $\alpha$ for which the contribution of the quartic in the velocity terms  is important in determining the end of inflation parameters.
For low  $\alpha < \alpha_{crit}$ the instantaneous temperature stays well below the bound  (\ref{tins3}) and is rather insensitive to changes with $\alpha$, for fixed values of $\xi, m$. This region belongs to the slow-roll regime and all data, including $T_{ins}$, can be derived  accurately enough employing the  the well-known  methodology based on the slow-roll functions. 
For high $\alpha > \alpha_{crit}$ slow-roll is not a valid approximation as we approach the end of inflation, and one should rely on a numerical treatment.

Lastly, in Figure \ref{figAsHiggs2},  we display 
the bound set on the instantaneous reheating temperature, and other observables,   for the Higgs model,  
for a  large value $\xi=10^5$.  The case displayed corresponds to  $m=3.05 \times 10^{-3}$ which is the lowest allowed by all data for this  $\xi$. 
This is actually fine-tuned since by slightly increasing the value of $m$ will move the $A_s$ predictions off its experimental bounds.  Note that this is a case where the entire allowed region falls  within the slow-roll regime.

Concluding this section, we state that for the Higgs model, 
as for the quadratic potential, $ V \sim h^2$ discussed in previous sections, and under the same conditions, 
 the bounds derived when reheat is instantaneous, hold for lower temperatures, as well.    
 In fact, for given $\xi$ the range of  the quartic coupling $\lambda $,  
 which we derived assuming instantaneous reheating, hold true for lower temperatures, as well.  The situation concerning bounds on the reheat temperature $T_{reh}$,  and on the parameter $\alpha$,  are similar to those of the minimally coupled model studied in the previous section, and thus need not be discussed in detail.
 However there is an important difference, regarding the bound $ r < 0.06$ which is much weaker in the non-minimal Higgs case ( $\xi \neq 0 $ ), allowing for much lower values of $\alpha$, as low as $ \alpha \simeq 10$, or even lower. Recall that in the case of the quadratic potential  $ V \sim h^2$, discussed in previous sections, $\alpha$ cannot be lower than about $ 10^{8}$, due mainly to the aforementioned  bound on the tensor to scalar ratio.
%%%%%%%%%%%%%% MINI          
%%%%%% MINI 

%%%%%%%%%%%%%%%%     
\begin{table}[H]
\begin{center}
\begin{tabular}{ccccc}  
  \hline \hline 
  \multicolumn{5}{c} {  Higgs Model  \quad ( pivot scale $ k^* = 0.05 \, Mpc^{-1}$ ) }    \\       
  \hline \hline   
    \multicolumn{5}{c} { Value of $ \quad  \xi = 0.1 $  } \\ 
  \hline
 min/max value of $\; \;  m $ : \quad   & \quad   \quad  & \quad $ m = 2.85  \times 10^{-6} $ \quad   & \quad &  \quad $ m = 3.05  \times 10^{-6} $ \\ 
     \hline
   \multicolumn{2}{l} {$\; \; \;  \quad \alpha \,$ } \quad  & $ 10 $ \quad \;  \; - \quad  \;  \; $ 10^9 $  &   &  $ \simeq 10^{13} $ \\
     \multicolumn{2}{l} {$\; \; \;  \; T_{ins}\,$ } \quad  & $  2.05  \times 10^{15} $ \quad  - \quad  $  1.97  \times 10^{15}$   &   &  $   3.82  \times 10^{14}$ \\
   \multicolumn{2}{l} {$\; \; \;  10^{9} \,A_s  \,$ } \quad  & $ 2.12 \; $  \quad -   \;  \quad $ 2.07 $  &   &  $ 2.08 $ \\
   \multicolumn{2}{l} {$\; \;  \; \; \; n_s  \,$ } \quad  & $ 0.9638 $ \quad  - \quad  $  0.9636 $  &   &  $  0.9609$ \\
   \multicolumn{2}{l} {$\; \; \;  \; \; r  \,$ } \quad  & $ 0.006 $ \quad  - \quad  $ 0.004 $  &   &  $ \sim 10^{-6} $ \\
   \multicolumn{2}{l} {$\; \; \;  \; N_* \,$ } \quad  & $ 56.11 \; $ \quad  - \quad  $ \; 55.86 $  &   &  $ 53.31$ \\
  % \hline
       \hline \hline
       \multicolumn{5}{c} { Value of $ \quad  \xi = 1.0 $  } \\  
  \hline
 min/max value of $\; \;  m $ : \quad   & \quad   \quad  & \quad $ m = 9.10  \times 10^{-6} $ \quad   & \quad &  \quad $ m = 9.70  \times 10^{-6} $ \\ 
     \hline
      \multicolumn{2}{l} {$\; \; \;  \quad \alpha \,$ } \quad  & $ 10 $ \quad \;  \; - \quad  \;  \; $ 10^{10} $  &   &  $ \simeq 10^{14} $ \\
     \multicolumn{2}{l} {$\; \; \;  \; T_{ins}\,$ } \quad  & $  1.70  \times 10^{15} $ \quad  - \quad  $  1.53  \times 10^{15}$   &   &  $   2.21  \times 10^{14}$ \\
   \multicolumn{2}{l} {$\; \; \;  10^{9} \,A_s  \,$ } \quad  & $ 2.10 \; $  \quad -   \;  \quad $ 2.07 $  &   &  $ 2.12 $ \\
   \multicolumn{2}{l} {$\; \;  \; \; \; n_s  \,$ } \quad  & $ 0.9632 $ \quad  - \quad  $  0.9630 $  &   &  $  0.9610$ \\
   \multicolumn{2}{l} {$\; \; \;  \; \; r  \,$ } \quad  & $ 0.0007 $ \quad  - \quad  $ 0.0004 $  &   &  $ \sim 10^{-7} $ \\
   \multicolumn{2}{l} {$\; \; \;  \; N_* \,$ } \quad  & $ 55.17 \; $ \quad  - \quad  $ \; 54.98 $  &   &  $ 52.71$ \\
\hline \hline
          \end{tabular}
\caption{Predictions  of the Higgs  Model ,  for $\xi=0.1$ and  $\xi=1.0$ , when $m$ gets its minimum and maximum allowed value allowed by all data, when reheating is instantaneous. The quartic coupling is $\lambda = m^2  $. 
These are actually  the cases displayed in Figure \ref{figAsHiggs}.  
For the lowest $m$ values allowed, a broad range of $\alpha$ is allowed, which is shrunk when $m$ gets its maximum value.
 The corresponding ranges for the instantaneous temperature and  the cosmological observables  
$ A_s, n_s, r,$,  as well as the number of efolds $N_* $,  are also shown corresponding to a pivot scale $ k^* = 0.05 \, Mpc^{-1}$.  
  }
\label{table111} 
\end{center}  
\end{table}  
%%%%%%%%%%%%%%%%%%%%%  

%%%%%%%%%%%%%% MINI        
%%%%%% MINI 

\noindent
{\bf{Summary of results -Discussion}}

\vspace*{3mm}

%%%%%%%
\noindent
The main  results reached, for the models  studied in this work, are  summarized in the following. 

For the Model I we have found that the allowed values for the mass parameter $m$ lie within rather tight limits
\bea
6.2 \times 10^{-6} \lesssim m \lesssim 6.8 \times 10^{-6} \, ,
\label{mrange}
\eea
while the  range of acceptable  $\alpha$ is 
\bea
10^8 \lesssim \alpha \lesssim 10^{14}  \, .
\label{manam}
\eea
%%%
Values of $\alpha < 10^7$ are not allowed, violating the bound set on $r$, and higher values  $\alpha > 10^{14}$ are incompatible with $n_s > 0.9607$. 
We remark that within the range (\ref{manam}), 
a critical value exists,   $\alpha_{crit} \simeq 10^{10}$, and for values of $\alpha$ beyond that  the contribution of the  quartic in the velocity terms are important in the determination of the end of inflation parameters. This is beyond the slow - roll regime. For such values of $\alpha$ the speed of sound approaches $ c_s^2 \simeq 2/3 $, its lowest mathematical bound in this kind of theories, and the upper bound set on the instantaneous temperature is saturated. 

For any given  $m$, within the range (\ref{mrange}), there exists a  range of $\alpha$-values, within (\ref{manam}), which  are compatible with  cosmological data. Assuming instantaneous reheating, 
and when $m$ gets its lowest value, in range (\ref{mrange}),  only values $\alpha \simeq 10^8$ are allowed.  In this case we have the largest possible  instantaneous temperature  $T_{ins} = 2.30 \times 10^{15} \, GeV$.  For this case $ r \simeq 0.05$, that is close to its upper  bound, and  the spectral index is $ n_s \simeq 0.9648$, that is it approaches its highest observational value. 
On the other hand, when $m$ receives its largest value, within  the range (\ref{mrange}), the allowed 
$\alpha$ are near the upper limit of (\ref{manam}),   $\alpha \simeq 10^{14}$. In this case we have the  lowest possible  instantaneous temperature  $T_{ins} \simeq 2.23 \times 10^{14} \, GeV$ while the value of $r$ is very tiny $ r \simeq 3 \times 10^{-8}$, and $n_s \simeq 0.9610$, that is close to its lowest observational bound.

Concerning the value of the tensor to scalar ratio $r$,  this model  is known to be in tension with $r$, predicting unacceptably large values  in metric formulation. However the situation  is rescued in the Palatini formulation  thanks to the appearance of  $\alpha \,  {\cal{R}}^2$  terms in the  action.

For the Higgs model, an additional parameter $\xi$ exists, which controls the non-minimal coupling of inflaton to gravity. Given the parameter $\xi$, bounds are set on the quartic Higgs coupling $\lambda$, defined in  (\ref{monmod2}), in the same way as for the Model I. 
For the benchmark values $ \xi = 0.1, 1.0 $ and $\xi =10^5$, considered in this work,  the allowed ranges for the parameter $m$, where $\lambda = m^2$, are as follows
\bea
\xi = 0.1 \quad & : & \quad 2.85 \times 10^{-6} \lesssim m \lesssim 3.05 \times 10^{-6} \nonumber \\
\label{mrange22}
\xi   = 1.0 \quad & :& \quad 9.10 \times 10^{-6} \lesssim m \lesssim 9.70 \times 10^{-6} \\
  \xi = 10^5 \quad & : & \quad \quad \simeq 3.05 \times 10^{-3}
\nonumber 
\eea
%%%
while the  range of acceptable  $\alpha$ is 
\bea
 \alpha \lesssim 10^{14}  \, .
\label{manam2}
\eea 
%%%
Values of   $\alpha > 10^{14}$ are incompatible with $n_s > 0.9607$. Note that there is no lower bound  on $\alpha$ arising from 
$ r < 0.06$, that is the $r$-bound weakens  in the Higgs case.  For the  case of Higgs,  
a critical value exists, as well, given by   $\alpha_{crit} = \frac{5}{3} \, ( \frac{\xi}{m})^2 $, and for values of $\alpha$ larger than $\alpha_{crit}$  the contribution of the  quartic in the velocity terms are important, as in the case of Model I.  In fact, for $\alpha > \alpha_{crit}$ the speed of sound approaches $ c_s^2 \simeq 2/3 $,  and the upper bound set on the instantaneous temperature is saturated. Note however that $ \alpha_{crit} $ exceeds the upper limit  (\ref{manam2}) for the case $\xi = 10^5$. 
This means that the contribution of the quartic in the velocity terms are negligible for  values of $\alpha$ that are of physicsl interest to us, for this value of $\xi$.

For any $m$ within (\ref{mrange22}),   ranges of $\alpha$-values exist that  are compatible with  the cosmological data.
For $\xi = 0.1$,  the range of allowed $\alpha$ is $ 10 - 10^9 $, when $m$ gets its lowest value $m =2.85  \times 10^{-6}  $, with ranges of instantaneous temperature  $T_{ins} = 2.05 \times 10^{15} - 1.97 \times 10^{14} \, GeV $. The value 
$ T_{ins} = 2.05 \times 10^{15}  \, GeV $ is the largest possible. 
The tensor to scalar ratio $r$ is in the range $r \simeq 0.006 - 0.004$, the highest values obtained in the Higgs model, while 
$n_s \simeq 0.9638 - 0.9636$.
For the largest  value $m =3.05 \times 10^{-6}  $ we can only have 
$\alpha \simeq 10^{13}$ and in this case $ T_{ins} = 3.82 \times 10^{14}  \, GeV $, while $ r \simeq 8.0 \times 10^{-7}$, which is pretty small, and $n_s \simeq 0.9609$.

Passing to $\xi = 1.0$ case, for the smallest $m$, $m =9.10  \times 10^{-6}  $, the allowed range of $\alpha$ is  $10 - 10^{10}  $ with $T_{ins} = 1.70 \times 10^{15} - 1.53 \times 10^{15} \, GeV $ and values of $r$ in the range $ r \simeq 7.0 \times 10^{-4} - 4.0 \times 10^{-4}$.  As for the spectar index $n_s \simeq 0.9632 - 0.9630$.
For the largest allowed $m$,   $m =9.70  \times 10^{-6}  $ the $\alpha$ parameter is around  $\alpha \simeq 10^{14}$, and in this case 
$ T_{ins} = 2.21 \times 10^{14}  \, GeV $, while $r$ is very low $ r \simeq 8.0 \times 10^{-8}$, and $n_s \simeq 0.9610$

 Finally for the case $\xi = 10^5$, $m$ is almost fine tuned to $m \simeq 3.05 \times 10^{-3}$. The allowed $\alpha$ - values span the region  $ \alpha \simeq 10 - 10^{14}$, with $T_{ins} = 1.23 \times 10^{14} - 1.21 \times 10^{13} \, GeV $, while $r$ is extremely low $ r \simeq7.5 \times 10^{-9}$ and $n_s \simeq 0.9608$.

Before leaving this section, we shall present a brief account of the metric and Palatini formulations of gravity. 
In Palatini gravity the affine connection $\Gamma$ is an  independent variable  and it is through the equations of motion that is connected to the metric $g_{\mu \nu}$.  In the absence of higher-$R$ terms in the action, and if additional scalars are present, that are minimally coupled to gravity the affinity  $\Gamma$ becomes the well-known Levi-Civita connection $ \Gamma = \Gamma(g)$ ( Christoffel symbols ). In this case the two theories yield identical results. However when non-minimal couplings exist and/or higher in the curvature $R$ terms are present the two theories differ.  

In the present work we have considered ${\cal{R}}^2$-terms coupled to  Palatini gravity as $\alpha \,  {\cal{R}}^2$ with $\alpha$ a constant. In the framework of metric gravity this theory describes, besides gravitons, the dynamics of  a scalar propagating degree of freedom, the scalaron with mass $ \sim 1/\sqrt{\alpha} $. This is best seen in the Einstein frame where the scalaron appears as a scalar field, which plays the role of the  inflaton,  moving under the influence of  the Starobinsky potential, which is predicted and not put in by hand.  In the framework of Palatini gravity the inclusion of $\alpha \,  {\cal{R}}^2$ terms differs  from the Starobinsky case of metric formulation. There are no extra propagating degrees of freedom and the would be inflaton, as well as its potential,  have to be introduced explicitly.  Note that the inclusion of additional scalars in the metric formulation, when   
$\alpha \,  {\cal{R}}^2$ is present, leads to multifield inflation in the Einstein frame.  

The inclusion of $\alpha \,  {\cal{R}}^2$ in Palatini  action has two important consequences.  It flattens the scalar potential in the Einstein frame, even for a steep original potential $V(h)$,  and a plateau is created, for large field values, which can sustain inflation. 
Besides, it induces quadratic terms in the velocity which affect the inflaton evolution towards the end of inflation for large  values of the coupling $\alpha$, larger than some critical value.
Both of these features are absent in the limit of $\alpha$ tending to zero, that is for sufficiently small couplings,  and thus are expected to play little role in the small $\alpha$-regime. 
Note that  
 the absence of $\alpha \,  {\cal{R}}^2$  terms does not imply that metric and Palatini formulations lead to  equivalent theories. 
 In fact, if  non-minimal couplings are present, as in the Higgs case for instance, the two formalisms lead to different actions in the Einstein frame and thus to different inflationary models.

On the phenomenological side, the inclusion of  $\alpha \,  {\cal{R}}^2$  terms  is well known to lower the value of $r$ for any inflationary model \cite{Antoniadis:2018ywb,Antoniadis:2018yfq,Enckell:2018hmo} .
The quadratic potential $\sim h^2$ belongs to this class which given unacceptably values for $r$ in metric formulation but it survives in the Palatini gravity, as we have seen.  Also in the Higgs case the values of $r$ are systematically lower than the Higgs inflation in metric formulation. Also the values of the non-minimal coupling  $\xi$ can take much  lower values  in the Palatini formulation.

%%%%%%%%%%%%%%%%%%%
\section{Conclusions}

We have studied models of ${\cal{R}}^2$-inflation in the framework of Palatini Gravity, where  the coupling of the 
${\cal{R}}^2$ to gravity is of the form $\sim \alpha \, {\cal{R}}^2$,  with  $\alpha$ is constant.  
The appearance of terms which are quartic in the velocity of inflaton is unavoidable in these theories. These  play little role, as being very small, during first horizon crossing, however they may play an important role in determining the end of inflation dynamics, and in particular the instantaneous reheating temperature, when the  scale $\alpha$ is large.  We have found that there is some critical value of $\alpha$ below which the mechanism of inflation follows slow-roll during the whole inflation era. However above this, the inflationary evolution  deviates from  slow-roll, as inflaton approaches the end of inflation, defined by where the acceleration 
$\ddot{a}$ of the cosmic scale factor vanishes.  In these case the determination of  end of inflaton is inaccurate when  slow-roll is employed.
 In a class of popular models, 
the speed of sound is bounded by   $ c_s^2 \geq  2 / 3$ putting upper bounds on the instantaneous reheating temperature $T_{ins}$, 
given by $ T_{ins} \leq   { 0.290 \,   m_{Planck}} / {\,  \alpha^{1/4}} $.  These  bounds are saturated for large values of $\alpha$. 

Assuming instantaneous reheating,  we have derived bounds on the parameters of the models studied in this work, arising  from the  amplitude of the scalar power spectrum $A_s$, the spectral index $n_s$ and the tensor to scalar ratio $r$.  The instantaneous 
reheating temperature cannot be arbitrarily small since   the  scale $\alpha$ is bounded from above by observations, imposing in turn lower bounds  on the inflationary scale.
For the models considered it is found that $\alpha$ can not exceed  $\alpha \simeq 10^{14}$, arising from $n_s$-data, which for  the largest $\alpha$ touches its lowest observational limit. 

When reheating is not instantaneous predictions  depend on the value of the effective equation of state parameter $w$. 
However, the constraints on the coupling  of the scalar potential   remain the same. 
For reasonable values of $w$, in the range $ w \simeq 0.0 - 0.25$,  the allowed values of $\alpha$ lie in a  certain range, 
constraining the  reheat temperature   $T_{reh}$  to be  within a range dictated by the value of $\alpha$ taken. 
%%%
When the  coupling of the scalar potential gets its minimum allowed value we obtain the highest possible temperature. 
This is  independent of the value of $w$ taken.  In fact, this is the instantaneous temperature corresponding to the smallest  allowed value of  $\alpha$, which receives its lowest possible value allowed by all data in this case. Then temperatures as large as 
$ \sim 10^{15} \, GeV$ can be reached, in principle,  for the models studied in this work. 
Low values  can be also obtained,  as low as   $ \simeq MeV$,   when the equation of state parameter is $ w \simeq 0.25$, or higher. The acceptable  values of $\alpha$ in this case are larger than $ \sim 10^{10}$, within the regime where slow-roll is not applicable. In this case  need go beyond slow-roll for a reliable cosmological study.

  \vspace*{4mm}  
{\textbf{Acknowledgments}}   
 A.B.L.  wishes to thank I. D.Gialamas for discussions at the initial stages of this work. He also thanks V. C. Spanos and 
 K. Tamvakis  for illuminating discussions.
%  \vspace*{7mm}  
% \newpage 

%%%%%%%%%%%%%%%%%%%%%%%%%  
%% \vspace*{0.5cm}

\end{document}